# High-throughput combinatorial approach expedites the synthesis of a lead-free relaxor ferroelectric system

*Di Zhang[1]*, Katherine J. Harmon[2], Michael J. Zachman[3], Ping Lu[4], Doyun Kim[5], Zhan Zhang[6], Nickolas Cucciniello[1], Reid Markland[1], Ken William Ssennyimba[1], Hua Zhou[6], Yue Cao[2], Matthew Brahlek[7], Hao Zheng[2,6], Matthew M. Schneider[8], Alessandro R. Mazza[1,8], Zach Hughes[1], Chase Somodi[1], Benjamin Freiman[1], Sarah Pooley[1], Sundar Kunwar[1], Pinku Roy[1], Qing Tu[5], Rodney J. McCabe[8], Aiping Chen[1]**

[1] Dr. D. Zhang, N. Cucciniello, R. Markland, K.W. Ssennyimba, Dr. A. R. Mazza, Z. Hughes, C. Somodi, B. Freiman, S. Pooley, Dr. S. Kunwar, Dr. P. Roy, Dr. A. P. Chen
Center for Integrated Nanotechnologies
Los Alamos National Laboratory
Los Alamos, NM 87545, USA
Email: dizhang@lanl.gov, apchen@lanl.gov

[2] Dr. K. J. Harmon, Dr. Y. Cao, Dr. H. Zheng
Materials Science Division
Argonne National Laboratory
Lemont, IL 60439, USA

[3] Dr. M. J. Zachman
Center for Nanophase Materials Sciences
Oak Ridge National Laboratory
Oak Ridge, TN 37830, USA

[4] Dr. P. Lu
Sandia National Laboratories
Albuquerque, NM 87185, USA

[5] Dr. D. Kim, Dr. Q. Tu
Department of Materials Science and Engineering
Texas A&M University
College Station, TX 77840, USA

[6] Dr. Z. Zhang, Dr. H. Zhou, Dr. H. Zheng
Advanced Photon Source (APS)
Argonne National Laboratory
Lemont, IL 60439, USA

[7] Dr. M. Brahlek
Materials Science and Technology Division
Oak Ridge National Laboratory

Oak Ridge, TN 37830, USA

[8] M. M. Schneider, Dr. A. R. Mazza, Dr. R. J. McCabe
Materials Science and Technology Division
Los Alamos National Laboratory
Los Alamos, NM 87545, USA

## Abstract

Developing novel lead-free ferroelectric materials is crucial for next-generation microelectronic technologies that are energy efficient and environment friendly. However, materials discovery and property optimization are typically time-consuming due to the limited throughput of traditional synthesis methods. In this work, we use a high-throughput combinatorial synthesis approach to fabricate lead-free ferroelectric superlattices and solid solutions of $(Ba_{0.7}Ca_{0.3})TiO_3$ (BCT) and $Ba(Zr_{0.2}Ti_{0.8})O_3$ (BZT) phases with continuous variation of composition and layer thickness. High-resolution X-ray diffraction (XRD) and analytical scanning transmission electron microscopy (STEM) demonstrate high film quality and well-controlled compositional gradients. Ferroelectric and dielectric property measurements identify the "optimal property point" achieved at the morphotropic phase boundary (MPB) with a composition of 48BZT-52BCT. Displacement vector maps reveal that ferroelectric domain sizes are tunable by varying $\{BCT\text{-}BZT\}_N$ superlattice geometry. This high-throughput synthesis approach can be applied to many other material systems to expedite new materials discovery and properties optimization, allowing for the exploration of a large area of phase space within a single growth.

## Introduction

Traditional lead-based ferroelectrics such as lead zirconate titanate $Pb(Zr_xTi_{1-x})O_3$ (PZT), though exhibiting outstanding piezoelectric properties, are not viable in many applications as processes including lead (Pb) have a negative effect on the environment and human health. Hence, lead-free ferroelectrics have been the subject of considerable research interest in recent decades. Tremendous effort has been devoted to improving the piezoelectricity, energy storage density and efficiency of lead-free dielectric systems (*1-3*). $BaTiO_3$ (BTO)-based perovskite ferroelectric oxides are one of the most studied and widely used lead-free ferroelectric systems, owing to their non-toxicity, simple fabrication process, and thermal and mechanical stabilities (*4, 5*). By substituting $Ba^{2+}$ (A site) or $Ti^{4+}$ (B site) ions in $BaTiO_3$ with other transition metal ions to form a variety of stable compositions including $(Ba_{1-x}M_x)TiO_3$ ($M$ = Na, K, Ca, Sr, *etc.*) and $Ba(N_yTi_{1-y})O_3$ ($N$ = Zr, Sn, Nb, Mn, *etc.*), the ferroelectric properties such as polarization, domain configuration, and coercivity can be remarkably tuned in these solid state solutions (*6-9*). In particular, the BTO-based relaxor ferroelectrics have been reported to show ultra-high piezoelectricity and efficiency of energy storage and conversion at the respective morphotropic phase boundary (MPB) compositions, making them ideal candidates for applications in non-volatile memory, neuromorphic computing, and energy storage devices (*10-12*). Consequently, the effect of various dopants on the ferroelectric properties of BTO-based systems has been intensively studied in recent years (*13-15*). However, specimens produced by conventional synthesis methods often contain fixed dopant levels, compositions, and geometries as reported in thin films, heterostructures, superlattices, and nanocomposites (*16*), making materials discovery across large chemical and geometric spaces challenging. Furthermore, unavoidable sample-to-sample variation introduces uncertainties into measured properties, which lessens the accuracy of systematic studies and reduces data reliability in material informatics.

High-throughput combinatorial synthesis approaches can accelerate new materials discovery by facilitating efficient identification of optimal functional properties (*17-19*). This is accomplished by rapidly surveying a large compositional landscape via synthesizing samples with a continuous compositional spread in a single experimental growth. In addition, the continuous variation of the local chemistry or geometry allows for a finer tuning of the composition as compared to conventional mix-and-measure approaches. Subsequent location-dependent structural characterization and property measurements then enable the creation of a "structure-property correlation map," which allows identifying the composition of a region with optimal properties. This high-throughput synthesis concept has enabled rapid discovery of new optical, electrical, magnetic, and structural materials (*20*-*23*). Using such a technique to expedite the discovery of relaxor ferroelectric thin films for energy storage is essential as there are large unexplored elemental and geometric spaces for these materials systems.

In this work, we use a high-throughput combinatorial pulsed laser deposition (cPLD) approach to expedite the exploration of film geometry and composition effects on ferroelectric properties in lead-free $(Ba_{0.7}Ca_{0.3})TiO_3$ (BCT) and $Ba(Zr_{0.2}Ti_{0.8})O_3$ (BZT)-based superlattices and thin films. To investigate the effect of SL geometry (*e.g.*, number of layers and layer thicknesses) and composition on the ferroelectric properties, we grow three series of BCT-BZT SL films with different numbers of layers while keeping the total film thicknesses nearly constant. High resolution X-ray diffraction (XRD) and analytical scanning transmission electron microscopy (STEM) experiments are conducted to reveal the epitaxial growth quality and microstructure of the BCT-BZT films. By systematically measuring ferroelectric properties of all samples, we establish "structure-property library maps" across different film geometries. By analyzing the polarization displacement vector maps, we further show that both the composition ratio and the geometry-controlled polar nanodomain size play critical roles in determining the properties of these films.

# Results

## Growth of BCT-BZT films

As illustrated in Figures 1A-B, BCT and BZT targets (see methods for details) are used in the thin film deposition. Five identically sized (5×10 $mm^2$) STO (001) substrates coated with ~ 60 nm thick $SrRuO_3$ (SRO) buffer layer, named from "*A*" to "*E*", are placed side by side on the substrate plate. During deposition, the substrate plate is kept static while the targets are rotating about their axes. The films studied here are grown by the following process. In "Phase 1" (Figure 1A), the BZT target is placed on-axis with substrate *A* so that the growth rate on substrate *A* (*E*) is the highest (lowest), which results from the natural angular distribution of the plume in PLD. This leads to a thickness gradient from substrate *A* (thickest) to *E* (thinnest). In "Phase 2" (Figure 1B), both the substrate plate and the target position are rotated such that substrate *E* is placed on-axis with the BCT target. In this case, the thickness gradient of BCT is opposite to that of the BZT deposited in Phase 1 such that the BCT layer on substrate *E* (*A*) is the thickest (thinnest). By alternating repeatedly between "Phases 1 & 2" during the deposition process, thin films with either a BZT-BCT superlattice (SL) structure or a solid solution structure are created. These films have a location-dependent SL geometry (*e.g.*, periodic BCT-BZT layers) and/or composition gradient. Therefore, both geometry and composition tuning are achieved via a simple process.

Figure 1C shows schematic cross-sectional views of the three $\{x\text{BCT-}(1\text{-}x)\text{BZT}\}_N$ ($N$ = 40, 80, 240) thin film series grown using the cPLD technique. Each series has 5 samples, labeled *A* to *E*, yielding a total of 15 samples across the three series. $x$ represents the composition of BCT, which is location dependent. In other words, $x$ changes continuously from *A* to *E*. Samples are defined herein by the letter 'S' followed by the number of layers *N* and the letter *A*-*E* of the sample within its series. For example, S40 refers to the series of 5 samples with alternately grown BCT and BZT with a total of 40 layers, and S40*C* refers to the specific sample *C* in the S40 series. As the total number of layers increase, the thickness of each layer is expected to reduce from ~12 unit-cells for S40*C* to ~2 unit-cells for S240*C*.

**Structural characterization of BCT-BZT films**

The as-grown BCT-BZT films were first measured at the Advanced Photon Source to characterize the film structure with a spatial resolution of a few tens of micrometers. The sample arrangement and X-ray scan direction is illustrated in Fig. S1. Fig. 1D displays the XRD intensities for S240. As shown, the STO and SRO peaks are constant, while the BCT-BZT (002) peaks exhibit a smooth shift from *A* to *E*, indicating a composition gradient. Fig. 1E shows the out-of-plane *c*-lattice parameters (*d*-spacing) for BCT-BZT, SRO, and STO as a function of location on the film. The *c*-lattice constant of S240 decreases from 4.05 Å at S240*A* (BZT phase dominant region) to 3.98 Å at S240*E* (BCT phase dominant region). This is consistent with the deposition setup that *A* is on-axis with the BZT target while *E* is on-axis with the BCT target. The lattice parameter difference can be explained by the cation radii relationship. Since $R_{Zr^{4+}}$ (0.72 Å) > $R_{Ti^{4+}}$ (0.605 Å) and $R_{Ca^{2+}}$ (1.00 Å) < $R_{Ba^{2+}}$ (1.35 Å), substituting $Zr^{4+}$ for $Ti^{4+}$ results in a larger lattice parameter in the BZT; in contrast, the substitution of $Ba^{2+}$ by $Ca^{2+}$ leads to a smaller lattice parameter for the BCT phase. The calculated *d*-spacing results from XRD are in good agreement with those previously reported for BCT-BZT films (*9*, *24*, *25*). It should be noted that the lattice parameter variation for the three samples in the middle (*e.g.*, *B*, *C* and *D*) seems linear.

Figs. 1F and 1G show the XRD data and *c*-lattice parameters of the S40 series. The first and second order satellite peaks are clearly seen in S40, indicating the periodic SL structures. The *d*-spacing values calculated from the (002) peak positions show the same varying trend and almost identical values to those of S240. Based on the ± 1$^{st}$ order satellite peak positions, the periodicity of the S40

SL is calculated to be ~9.5 nm at S40*C*, and it decreases toward the edges (*A*, *E*), reaching ~8.5 nm at S40*E* (Fig. S2). This thickness profile is typical for the cPLD method and has been reported previously (*26*-*28*). Similar XRD results are observed for the S80 series (Fig. S3) with a SL periodicity of ~4.7 nm at S80*C* and ~4.2 nm at S80*A* or S80*E*.

The microstructures of these films were further characterized via STEM. The low-magnification high-angle annular dark-field (HAADF) STEM images shown in Fig. 2A-C for the S40*C*, S80*C* and S240*C* BCT-BZT films reveal that the S40*C* and S80*C* films exhibit obvious SL structures, whereas the S240*C* film shows a single phase, indicating the intermixing of BCT and BZT phases, as reflected in the schematics of Fig. 1. HAADF-STEM images (Fig. 2D-F) and energy-dispersive X-ray spectroscopy (EDS) maps of Ca and Zr (Fig. 2G-I) confirm the SL structures of the S40*C* and S80*C* films as well as the single-phase of S240*C*. Based on these images and maps, the periodicity of the S40*C* and S80*C* SLs, *i.e.* BCT/BZT bilayer thickness, are estimated to be ~10.0 and ~5.0 nm, which is in good agreement with the calculated values from the XRD data (Fig. S2 and Fig. S3D). If S240*C* were to form a SL structure, the periodicity would be ~ 1.7 nm, corresponding to 2 unit-cells of BZT and 2 unit-cells of BCT. However, the high-magnification HAADF-STEM image shown in Fig. 2F exhibits a uniform single-phase structure for S240*C* without any visible phase or grain boundaries. Due to a fast growth rate of 0.6-0.7 Å/pulse for the on-axis growth (near samples *A* or *E*), interdiffusion across different layers resulted in a solid solution for S240. Interdiffusion therefore provides a unique way to generate solid solution thin films with a composition gradient.

We also investigated the microstructures of S40 from *A* to *E* to confirm the composition and geometry gradient, as shown in Fig. S4. HAADF-STEM images and EDS elemental maps of Ca and Zr show a clear compositional variation from BZT dominant in S40*A* (~75% volume of BZT) to BCT dominant in S40*E* (~70% volume of BCT). Furthermore, geometric phase analysis (GPA) was performed on the HAADF-STEM along the out-of-plane ($\varepsilon_{yy}$) and in-plane ($\varepsilon_{xx}$) directions to directly visualize lattice variations in the BCT-BZT SLs, as shown in Fig. S5. The reference regions were taken from the clean and undistorted BZT interlayer areas in all the HAADF-STEM images. The out-of-plane ($\varepsilon_{yy}$) strain maps show abrupt changes at the BCT and BZT interfaces due to their different intrinsic lattice constants, which result in tensile (positive) strain in BZT and compressive (negative) strain in BCT. Moreover, it is observed that S40*C* film (Fig. S5B) exhibits a homogeneous in-plane ($\varepsilon_{xx}$) strain map, while the $\varepsilon_{xx}$ maps of S40*A* (Fig. S5D) and S40*E* (Fig. S5F) films show in-plane SL like structure, which might be related to some in-plane domain epitaxy structures formed during the film growth. The S80*C* film shows a typical BCT/BZT out-of-plane SL structure and in-plane SL like structure as well (Fig. S6B). More interestingly, the out-of-plane strain map of S240*A* also shows some SL like structures, but no SL structure was observed in the corresponding HAADF-STEM image (Fig. S6D). This can be ascribed to the extreme sensitivity of GPA strain analysis to atomic-scale lattice variations, while HAADF-STEM *Z*-contrast imaging is more sensitive to elemental character.

**Ferroelectric and dielectric properties of BCT-BZT films**

Next, ferroelectric properties of the BCT-BZT films were investigated. For simplicity, the results of S40 and S240 films are shown in Figure 3 for direct comparison. Fig. 3A and 3B show ferroelectric polarization (*P*) maps of the S40 and S240 films measured under 700 kV/cm electric field at three different frequencies of 1 kHz, 10 kHz, and 100 kHz. Six polarization datapoints were measured at each sample (see Fig. S1) along the composition gradient direction. Thus, there are 90 data points shown in each polarization map. The polarization values of both series are found to exhibit similar trends with the highest *P* values obtained near sample *D* (yellow shaded area).

At an electric field of 700 kV/cm, the highest *P* values for S40*D* and S240*D* are estimated by Lorentzian fitting to the data to be ~18 $\mu$C/cm$^2$ and ~22 $\mu$C/cm$^2$, respectively. Figure 3C and 3D show the energy storage density (*W*) and energy storage efficiency ($\eta$) of the S40 and S240 films calculated based on their measured polarization *vs.* electric field (*P-E*) hysteresis loops. For both S40 and S240, *W* first increases to a maximum near sample *D* and then decreases. The $\eta$ values for S40 follow the same trend, reaching a maximum value of ~80% at S40*D*. The $\eta$ values for S240 are location independent (~80-91% from *A* to *E*), with the maximum $\eta$ (91%) higher than the current state-of-the-art in PZT, BTO-BFO-STO, and BZT-BCT systems (*12, 29, 30*). In sum, *W* and $\eta$ reach their maximum values near sample *D* location. In addition, the electric field and frequency dependent data for S40 and S240 films (Fig. S8) demonstrate that the energy storage density *W* is noticeably impacted by the magnitude and frequency of the applied electric fields, while energy storage efficiency $\eta$ values in general are much more stable across different measurement parameters.

Piezoresponse force microscopy (PFM) was used to confirm the ferroelectricity and obtain the piezoelectric coefficient ($d_{33}$) values of the samples. More detailed information can be found in Fig. S9 and Supplementary Materials. The $d_{33}$ values of S40 and S240 are summarized in Figure 3E. Both series show the highest $d_{33}$ values near sample *D*, and the minor left-shift of the maximum values compared to those in the polarization maps (Figs. 3A and B) may be attributed to coarser sampling in the PFM measurements. It is also noted that the $d_{33}$ values of S240 films are, in general, much higher than those of S40 films. This indicates that the mixed solid solution phase of BZT-BCT films has higher $d_{33}$ values than BZT/BCT SL films, which is consistent with previous reports (*31-33*). Figure 3F depicts the schematic illustration of the PFM experiment setup and the "*Los Alamos National Laboratory*" logo written on the S240*D* film surface (close to the highest $d_{33}$ point at *D*) by controlling the local ferroelectric domains via electrical polarization. The clear piezoresponse amplitude contrast between the polarized and unpolarized regions confirms the ferroelectric response of the film. Furthermore, based on the *P-E* hysteresis loops of both S40*D* and S240*D* (Fig. S10), the S240 film (Fig. S10C) exhibits smaller coercivity than the S40 film (Fig. S10B). The slim *P-E* loops, small remnant polarizations and coercivity fields indicate a possible relaxor ferroelectric behavior for S240. This is consistent with our recent work on the growth of BZT-BCT solid solution films that show relaxor ferroelectric behavior in this system (*30, 34*). The coercive fields ($H_c$) of both films at samples *C* and *D* are further extracted and plotted in Fig. S11, showing that the lowest $H_c$ of both films are obtained near the center of sample *D*. Furthermore, fatigue and retention measurements (Fig. S12) show that both films exhibit near zero reduction of the remanent polarization for over $10^9$ cycles or 24 hours after the experiment was terminated, indicating a strong stability of the remanent polarization against cycling fatigue and retention time.

The dielectric properties of the S40 and S240 BCT-BZT films were obtained from the measured capacitance *vs.* voltage (*C-V*) curves. The permittivity values are found to follow the same location-dependent trend as polarization (Figs. 3A, B) and energy storage density (Figs. 3C, D). Namely, as shown in Figs. 3G and 3H, the dielectric permittivity ($\varepsilon_r$) of both series exhibits the highest $\varepsilon_r$ values at sample *D*, increasing from ~130 at sample *A* to ~300 at sample *D* in S40 and from ~320 at *A* to ~680 at *D* in S240. Overall, S240 exhibits higher permittivity and polarization across samples than S40. The insets in Fig 3G and 3H show typical butterfly-shaped *C-V* curves at sample *D,* demonstrating the ferroelectric nature of these films. The measured dielectric losses (tan $\delta$) of both films show limited location-dependence and are less than 0.5, indicating the stable dielectric properties for both series. Fig. 3I summarizes the dielectric tunability and loss tangent of both films at different sample locations. It is noted that S240 maintains a high dielectric tunability of 60~70% while S40 has lower tunability values of 35~45 %. More interestingly, the dielectric tunability is insensitive to location and is, therefore, likely controlled by film geometry (SL or single phase) rather than composition. Hence, in terms of properties including polarization, piezoelectric coefficient, permittivity, and dielectric tunability, the mixed solid solution BZT-BCT films (S240) are superior to SL films (S40).

**Property tuning mechanism of BCT-BZT films**

It is well known that for ferroelectric materials the high piezoelectricity stems from its morphotropic phase boundary (MPB), where the free energy is isotropic with very low energy barrier for polarization rotation and lattice distortion (*10*). Previous works show that the highest piezoelectric and dielectric properties are obtained at 50BZT-50BCT bulk ceramics (*10, 13*). Based on our measurements, the best ferroelectric and dielectric properties of the BCT-BZT thin films are present near sample *D*. Therefore, it is essential to investigate the microstructure and identify the BCT: BZT phase ratio for the SL and solid solution films at their "optimal property points". To obtain this critical information, we performed high-resolution analytical STEM, including imaging, EDS mapping and elemental quantification, and electron energy-loss spectroscopy (EELS) for the S40*D* and S240*D* at their "optimal property locations". Figs. 4A-C show the HAADF-STEM image and corresponding out-of-plane ($\varepsilon_{yy}$) and in-plane ($\varepsilon_{xx}$) strain maps of S40*D* film. It is interesting to note that the $\varepsilon_{yy}$ strain map shows a somewhat curved layer interface in contrast to the comparatively straight BCT/BZT interface observed in the HAADF image. This indicates that the strain field within the SL structure is more sensitive to the local defects or grain boundaries which are not visible in the *Z*-contrast HAADF image. The $\varepsilon_{xx}$ map shows the in-plane inhomogeneity of the SL film. The composite EDS elemental map for Ca and Zr (Fig. 4D), along with their intensity line profiles (Fig. 4E and Fig. S13), reveal the typical $(BCT/BZT)_N$ SL structure with each layer being ~5 nm thick. The distinct BCT/BZT interface indicates the limited interdiffusion during film growth. EDS elemental quantification for S40*D* is shown in Table S1, where the Ca/Zr atomic ratio is ~ 1.07. Meanwhile, the EDS elemental maps of the S240*D* solid-solution film are shown in Fig. S14. EDS elemental quantification results for S240*D* are shown in Table S2, and the Ca/Zr atomic ratio is ~ 1.10. Therefore, for both the S40 and the S240 film series, the "optimal property" is achieved for BCT: BZT $\approx$ 1.07-1.10, corresponding to a chemical composition of ~ 48BZT-52BCT. The minor difference between our MPB composition (48BZT-52BCT) and the reported MPB composition in the bulk BCT-BZT ceramics (50BZT-50BCT) can be attributed to the fact that thin films are often subjected to lattice strains, oxygen vacancies and other defects which are not present in bulk materials, which may lead to a shift of the triple point in the BCT-BZT phase diagram.

The above results show that the SL periodicity plays a critical role in impacting the ferroelectric and dielectric properties. Although the S240 series exhibits enhanced ferroelectric properties relative to the S40 series, S240 exhibits higher dielectric leakage (Fig. 3I). The dielectric loss in ferroelectrics is found to be closely related to defect chemistry and charge-trapping phenomena within the studied materials, where oxygen vacancies play an important role in these processes (*35-37*). Hence, it is necessary to investigate the oxygen vacancies concentration ($V_O$) and valence states of transitional metals for the BCT-BZT films near its MPB composition. We therefore conducted EELS experiments for S40*D* and S240*D*, as summarized in Fig. 5. The same experiment setting such as energy dispersion and dwell time were used for all the EELS data acquisition at multiple locations on each sample with similar thicknesses, thus the impact of those extrinsic factors on the measured results can be excluded. Figs. 5A and 5B display EELS O-*K* energy-loss near-edge structure (ELNES) extracted from different locations within the S40*D* SL and S240*D* solid-solution films and STO substrate (for reference), respectively. Several fine-structure features are present and labeled “A”, “B”, “C”, and “D”. The pre-peak “A” represents transitions from the O 1*s* to unoccupied O 2*p* hybridized with Ti 3*d* band states (*38-40*). Both S40*D* and S240*D* films show decreased pre-peak intensities compared to that of the STO substrates, indicating the decreased number of unoccupied states in Ti 3*d* band for the BCT-BZT films. This can be explained by the valence change of Ti from $Ti^{4+}$ to $Ti^{3+}$, which is accompanied by doping one electron per site into the empty Ti 3*d* band ($3d^0$ to $3d^1$), making the excitation of these states less likely and leading to a decrease in the pre-peak intensities. The “B”, “C”, and “D” peaks mainly originate from transitions of electrons from O 1*s* core states to O 2*p* states that are hybridized with *A*-site cation (*i.e.* Sr 4*d* and Ba 5*d*) bands, Ti 4*s* bands, and Ti 4*p* bands, respectively (*38,41-43*). It is noted that the “C” peak in both films splits into two subpeaks labeled as “$C_1$” and “$C_2$”. This can be explained by the displacement of Ti atoms in the BTO octahedra that leads to polarization breaking the symmetry of oxygen atoms and causes the energy splitting (*39*). Overall, the O *K*-edges of both BCT-BZT films exhibit damped intensities for all peaks, which can be explained by the presence of oxygen vacancies in the studied materials (*38,44-46*).

To further investigate the oxidation (valence) states of Ti, we collected EELS spectra of the Ti $L_{2,3}$ edges for the S40*D* and S240*D* films, as plotted in Fig. 5C and 5D. In comparison to the STO reference, the peak positions of the $L_{2,3}$ edges of both films undergo minor shifts to lower energy losses, possibly indicating a slight reduction of the Ti oxidation state (*46-48*). The $L_3/L_2$ intensity ratio can also be correlated with the oxidation states of middle or late 3*d* transition metals (TM) (*47-50*). However, for early 3*d* TM such as Ti, this method is difficult to apply due to the heavy overlap of the $L_2$ and $L_3$ edges. To address this, Stoyanov *et al.* and Shao *et al.* proposed calculating the Ti $L_{2,3}$ intensity ratio using a reduced energy window (*48,51*). An energy window of 1 eV is therefore used here to calculate the Ti $L_3/L_2$ intensity ratio, after removing the background portion of the raw Ti-$L_{2,3}$ ELNES spectra using a Hartree-Slater modeled *L* edge atomic cross-section step function (*47-49*). Figs. 5E and 5F summarize the calculated Ti $L_{2,3}$ ratios extracted from different locations on the S40*D* and S240*D* films, with error bars that capture the range of values measured in multiple spectra. It is shown that the Ti $L_{2,3}$ ratio in STO substrate is ~0.75. For the S40*D* SL, the $L_3/L_2$ ratio slightly increases to a value higher than 0.77 in both the BZT and BCT phase regions and to ~0.79 at the BZT/BCT interface; for the S240*D* solid solution, the $L_3/L_2$ ratio is higher at 0.79 and increases to 0.80 near the surface of the film. The multiple spectrum collections in the same area show consistent results with the standard deviation being less than 0.005. Considering that TM oxidation states decrease with an increase in the $L_3/L_2$ ratio (*47-51*) and that there are more noticeable $L_{2,3}$ peak shifts in the S240*D* film, we conclude that the average Ti valence state of the S240*D* film is lower than that of the S40*D* SL film. This indicates a higher $V_O$ concentration in the solid-solution film than in the SL film. This finding can also explain the higher dielectric loss of the S240*D* film (Fig. 3I), which further confirms that oxygen vacancies are a key factor impacting dielectric properties. The Ba-$M_{4,5}$ edges of both films show no obvious peak shift (Fig. S15), and no $M_{4,5}$ intensity changes are detected, indicating that the valence states of Ba atoms (*A*-site) are much less sensitive to the oxygen stoichiometry than the valence states of Ti (*B*-site) for the perovskite structure ($ABO_{3-\delta}$) oxides.

Recently, polar displacement vector maps based on atomic-resolution STEM imaging has been demonstrated as an effective method for visualizing the polarization fields and domain structures of ferroelectric materials (*52-55*). In this work, based on aberration-corrected *Z*-contrast HAADF-STEM images, we generate polarization displacement vector maps along the <100> projection axis for the S40, S80, and S240 films near their MPB compositions using TopoTEM (*56*) (see Supplementary Materials for details). As shown in Figure 6, the projected polarization vector of each Ti atom ($\delta_{Ti}$) is overlaid on the original STEM image, with the color of the $\delta_{Ti}$ vector determined by its polarization direction. From this displacement vector map we can identify the differently oriented polar nanodomains (PND). Figs. 6B and 6C show the displacement vector maps of the S40*D* and S240*D* films, respectively. The corresponding raw HAADF-STEM images are shown in Fig. S16A and S16B. The yellow dashed lines in Fig. 6B indicate the BCT/BZT interfaces of the SL film. The majority of Fig. 6B shows uniform polar vector directions, while differently oriented polar domains (shown in dashed colored circles) are present mainly near the BCT/BZT interface. In contrast, the vector map of the S240D solid solution film (Fig. 6C) exhibits many randomly oriented PNDs with smaller average sizes. The STEM image and corresponding displacement vector map of the S80C film (Fig. S16C and D) show a higher number of differently oriented PND than observed in the S40 SL (Fig. 6B), most of which are near the BCT/BZT interfaces. These polarization displacement maps of $\{BCT\text{-}BZT\}_N$ ($N$ = 40, 80, 240) films reveal that the PND sizes and distributions within ferroelectric thin films can be artificially controlled via tuning the thickness and spacing of polar layers. The decrease of ferroelectric interlayer thickness leads to smaller size and random orientations of PND and promotes the transition from regular ferroelectric-like SL to relaxor-like behavior of the solid solutions.

## Conclusions

We have demonstrated a high-throughput combinatorial pulsed laser deposition (cPLD) technique to grow epitaxial ferroelectric BCT-BZT films with superlattice structure and mixed solid solution phase. A compositional gradient is achieved across a 1-inch film area. By tailoring the thickness of each SL layer from several nanometers down to 2 unit cells, the film microstructure transitions from a distinct SL to single-phase solid solution, with an associated change from ferroelectric-like to relaxor-like behavior due to the tuning of polar nano domain (PND) size. Location-dependent ferroelectric and dielectric property maps enable selection of the "optimal property point" of the nanocomposite and identify its morphotropic phase boundary (MPB) composition point from a single specimen. This high-throughput combinatorial synthesis approach can be applied to many materials systems to significantly expedite the materials discovery and property optimization processes.

## Acknowledgements

**Funding:** The work at Los Alamos National Laboratory was supported by the NNSA's Laboratory Directed Research and Development Program, and was performed, in part, at the Center for Integrated Nanotechnologies, an Office of Science User Facility operated for the U.S. Department of Energy (DOE) Office of Science by Los Alamos National Laboratory (contract 89233218NCA000001) and Sandia National Laboratories (contract DE-NA0003525). Los Alamos National Laboratory, an affirmative action equal opportunity employer, is managed by Triad National Security, LLC for the U.S. Department of Energy's NNSA, under Contract No. 89233218CNA000001. Sandia National Laboratories is a multiprogram laboratory managed and operated by National Technology and Engineering Solutions of Sandia, LLC, a wholly owned subsidiary of Honeywell International, Inc., for the U.S. Department of Energy’s National Nuclear Security Administration under contract DE-NA0003525. This paper describes objective technical results and analysis. Any subjective views or opinions that might be expressed in the paper do not necessarily represent the views of the U.S. Department of Energy or the United States Government. The synchrotron X-ray diffraction measurements done at the Argonne National Laboratory was supported by the U.S. Department of Energy, Office of Science, Basic Energy Sciences, Materials Science and Engineering Division, and the XRD data analysis was based on work supported by Laboratory Directed Research and Development funding from Argonne National Laboratory, provided by the Director, Office of Science, of the U.S. Department of Energy and by the U.S. DOE Office of Science-Basic Energy Sciences under Contract No. DE-AC02-06CH11357. This research used resources of the Advanced Photon Source, a U.S. Department of Energy (DOE) Office of Science user facility operated for the DOE Office of Science by Argonne National Laboratory under Contract No. DE-AC02-06CH11357. The aberration-corrected HAADF-STEM portion of this research was supported by the Center for Nanophase Materials Sciences (CNMS), which is a US Department of Energy, Office of Science User Facility at Oak Ridge National Laboratory. The PFM measurements at Texas A&M University were supported by the donors of the ACS Petroleum Research Fund under Doctoral New Investigator Grant 62603-DNI10. Q.T. served as Principal Investigator on ACS 62603-DNI10 that provided support for D.K.

**Competing interests:** The authors declare no conflict of interest.

**Data and materials availability**

## Supplementary Materials

Materials and Methods

Fig. S1-S16

References (*57-62*)

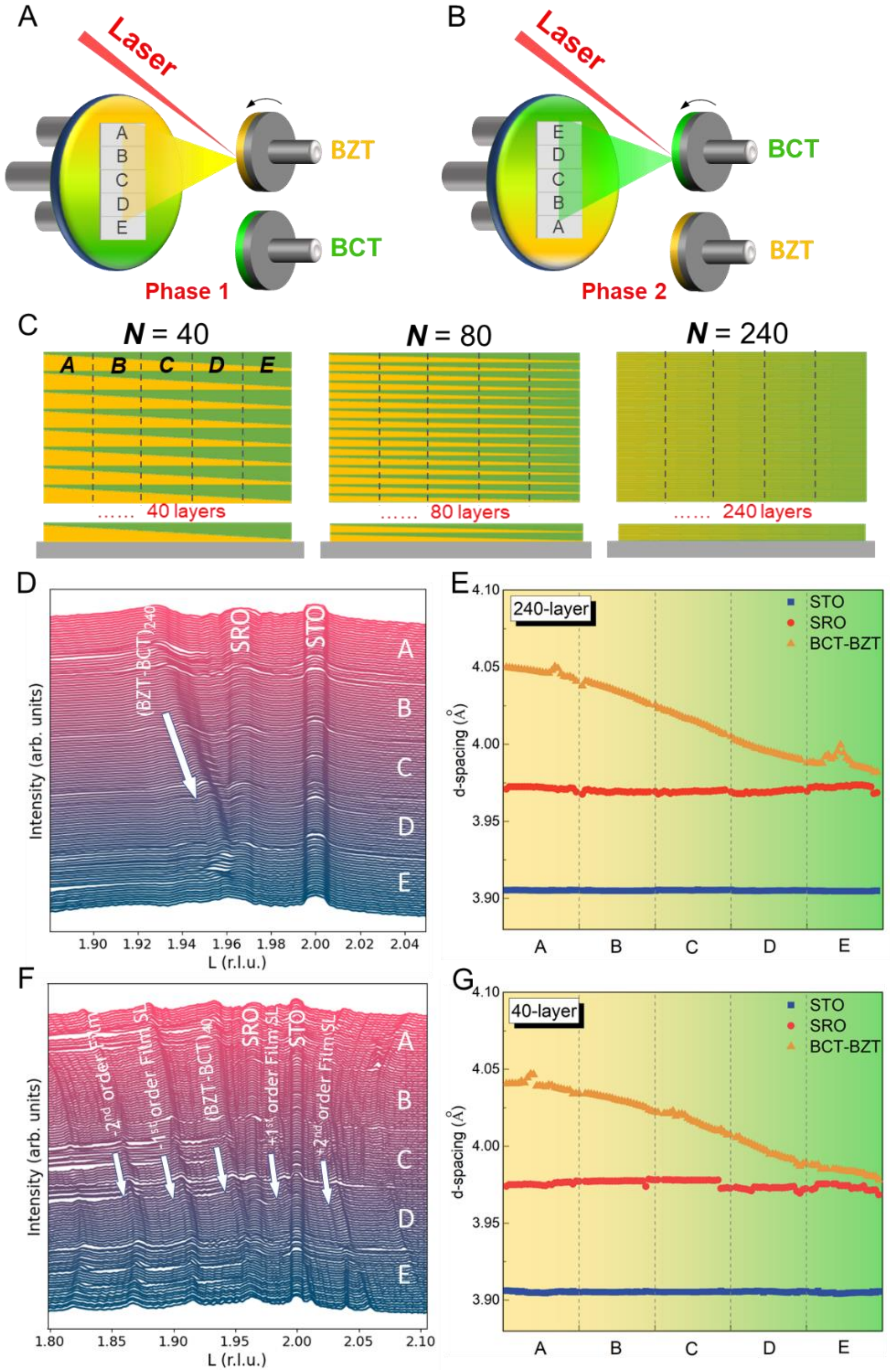

**Fig. 1 Combinatorial synthesis and XRD patterns of $x(Ba_{0.7}Ca_{0.3})TiO_3 – (1\text{-}x)Ba(Zr_{0.2}Ti_{0.8})O_3$ ($x$BCT-(1-$x$)BZT) thin films.** (**A**, **B**) Schematics of the combinatorial synthesis approach of growing the BCT-BZT thin films for Phase 1 and 2. (**C**) Schematics of the cross-sectional views of S40, S80, and S240 BCT-BZT films. XRD intensities and calculated *c*-lattice parameters (*d*-spacings) of (**D**, **E**) S240 and (**F**, **G**) S40 BCT-BZT films. The data were extracted from (002) reciprocal space maps. The film *d*-spacings as a function of position along the sample were evaluated from the peak L positions assuming Gaussian intensity distributions. r.l.u. = reciprocal lattice units. Note: the fringe patterns between two superlattice peaks are not real film thickness fringes but some artifacts due to the data processing of 3DRSM profiles.

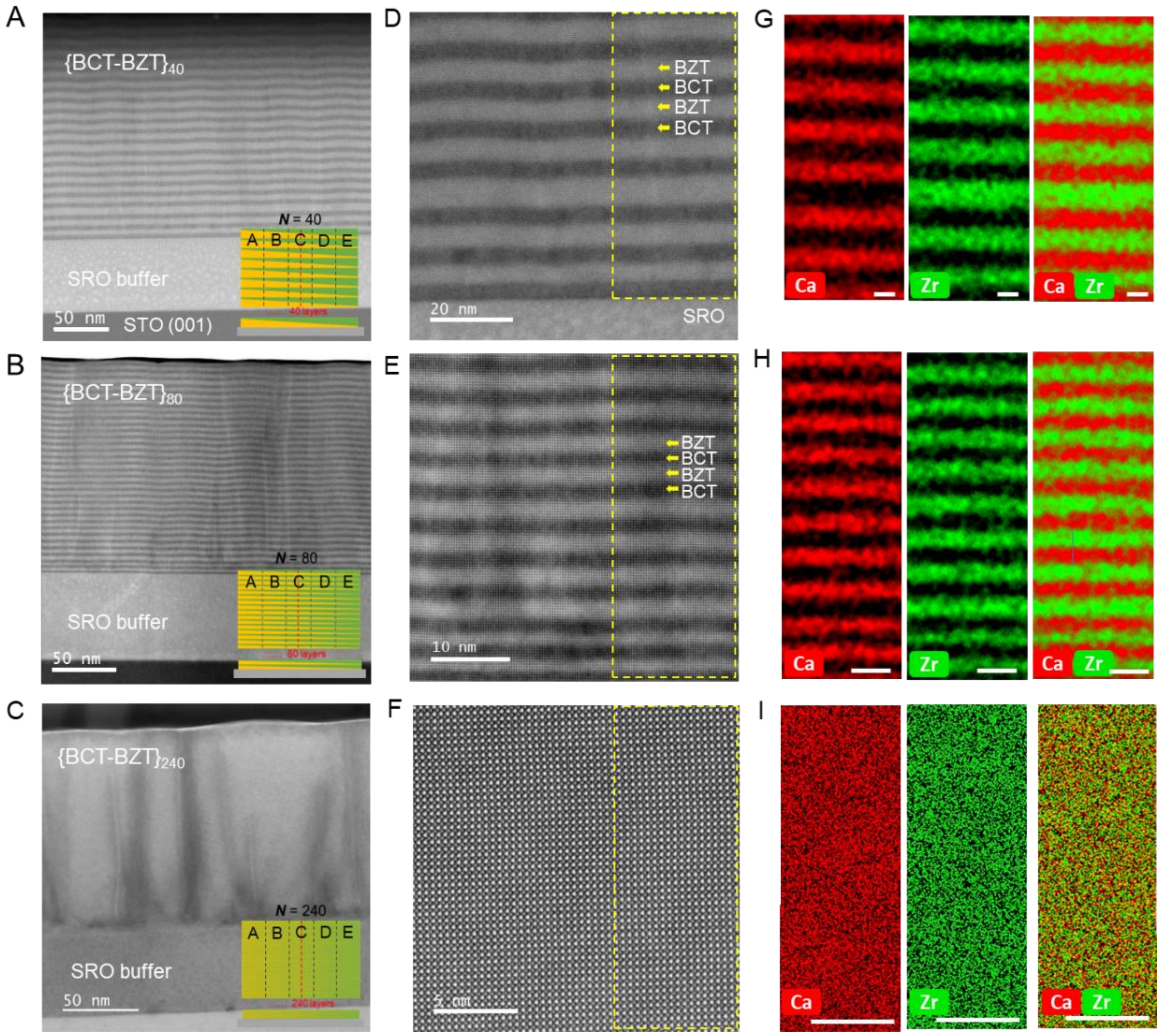

**Fig. 2 Microstructural characterization of BCT-BZT films.** (**A, B, C**) Low-magnification HAADF-STEM images of S40, S80, and S240 BCT-BZT thin films at center of location *C*. Insets are the schematics of the cross-sectional views of S40, S80, and S240 BCT-BZT films with the red dashed lines showing the investigated sample locations. (**D**, **E**, **F**) Atomic-resolution HAADF-STEM images of S40, S80, and S240 BCT-BZT thin films, where yellow dashed rectangles indicate areas of EDS maps. (**G**, **H**, **I**) Energy-dispersive X-ray spectroscopy (EDS) maps of Ca and Zr in S40, S80, and S240 BCT-BZT films. Scale bars in G, H, I: 5 nm.

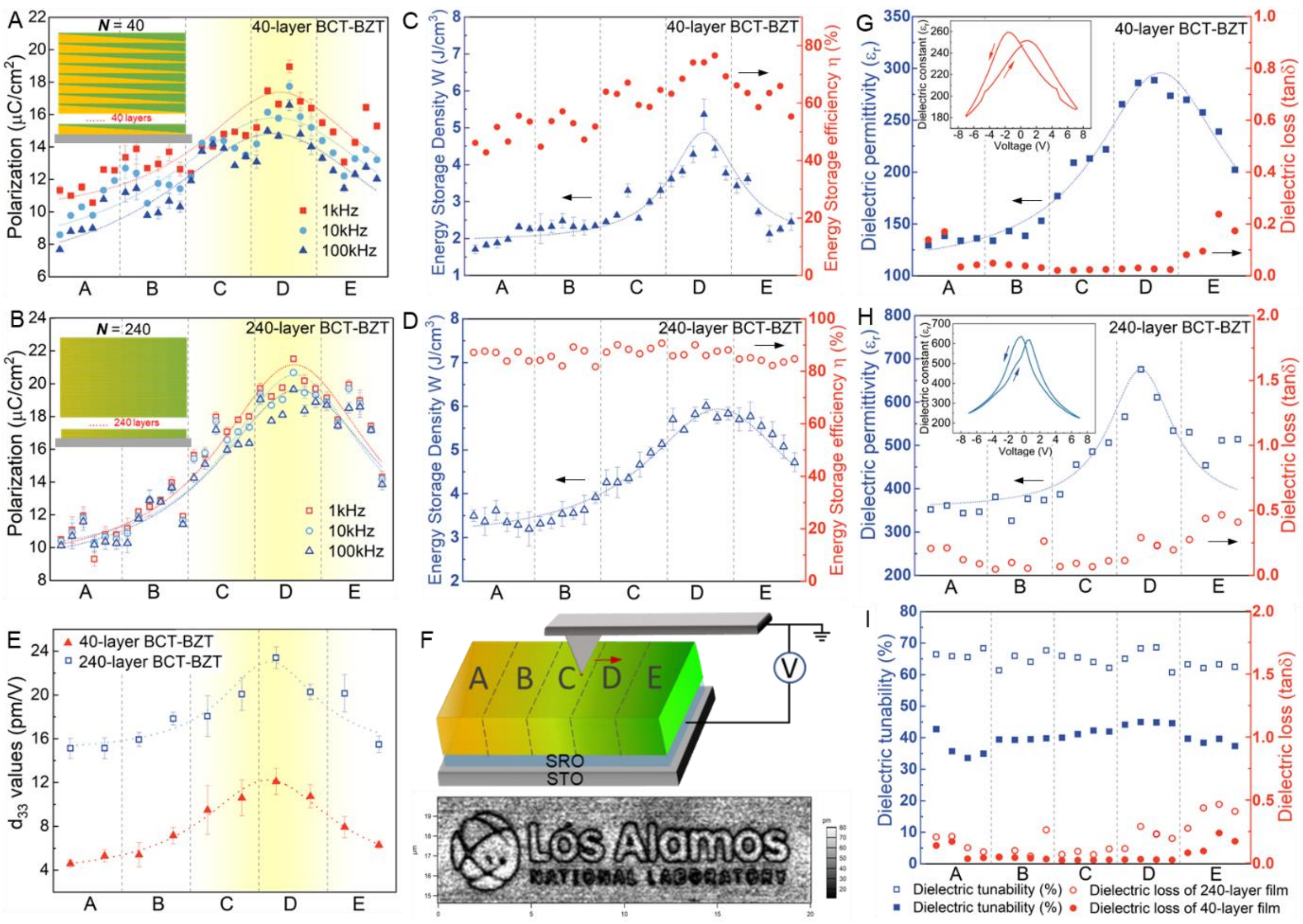

**Fig. 3 Ferroelectric and dielectric properties of the *x*BCT-(1-*x*)BZT thin films.** The location-dependent (**A**, **B**) ferroelectric polarization (*P*), (**C**, **D**) energy storage density (*W*) and efficiency ($\eta$), (**E**) piezoelectric coefficient ($d_{33}$), (**G**, **H**) dielectric permittivity ($\varepsilon_r$) and dielectric loss (tanδ), and (**I**) dielectric tunability (%) of S40 and S240 BCT-BZT thin films. (**F**) Schematic of the piezoelectric force microscopy (PFM) measurement and the vertical PFM amplitude map showing the “Los Alamos National Laboratory” logo. Insets are the schematic cross-sectional views of (A) S40 and (B) S240 BCT-BZT films; measured capacitance *vs.* voltage (*C*-*V*) curves of (G) S40 and (H) S240 BCT-BZT films.

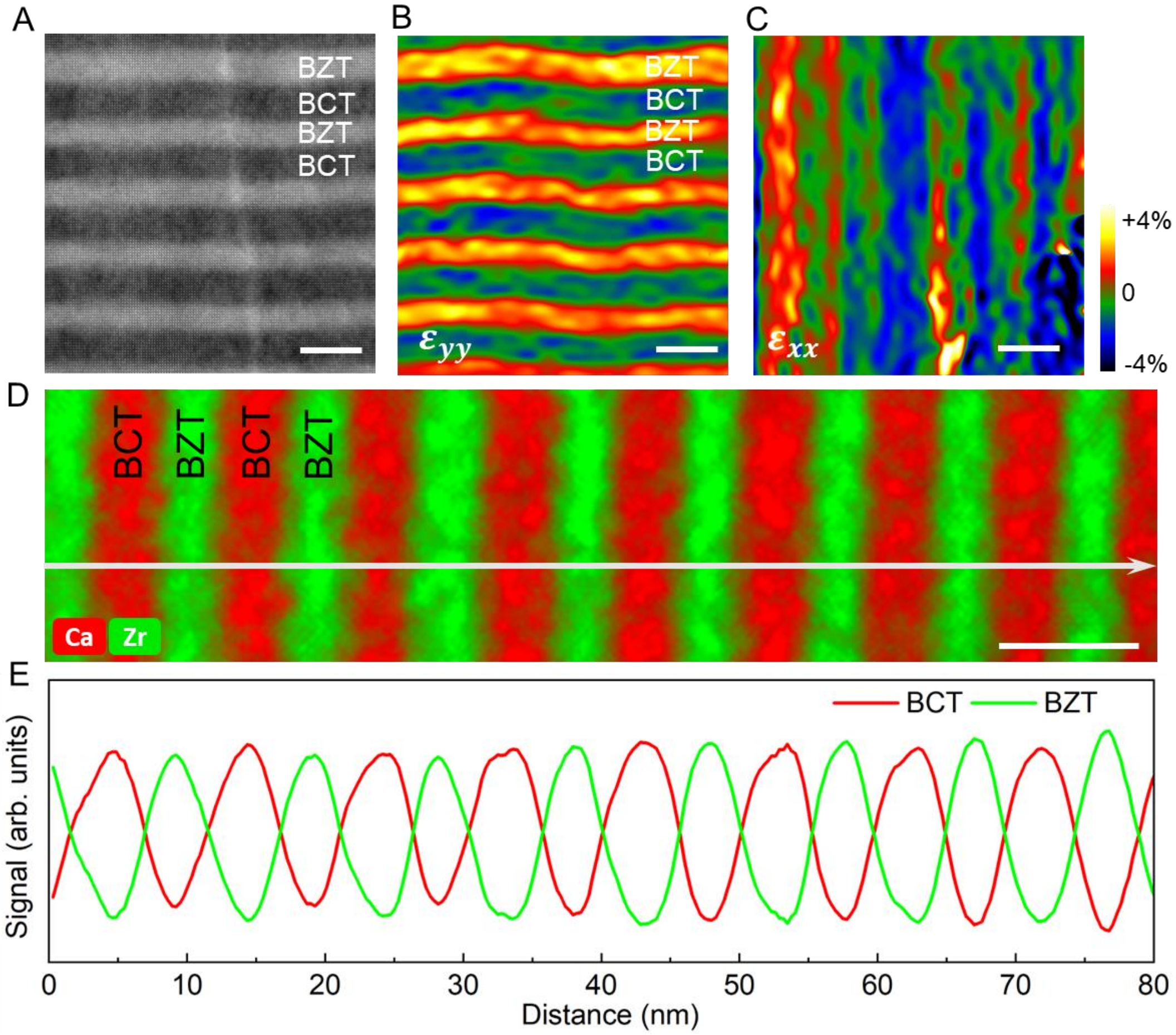

**Fig. 4 Microstructure analysis of S40*D* BCT-BZT film at the "optimal property point".** (**A**) HAADF-STEM image and (**B**, **C**) corresponding GPA analysis of out-of-plane ($\varepsilon_{yy}$) and in-plane ($\varepsilon_{xx}$) normal strain. (**D**) EDS elemental composite map of Ca (red) and Zr (green) and (**E**) elemental intensity line profile (raw counts) along the white arrow direction in (D). Scale bars: 10 nm.

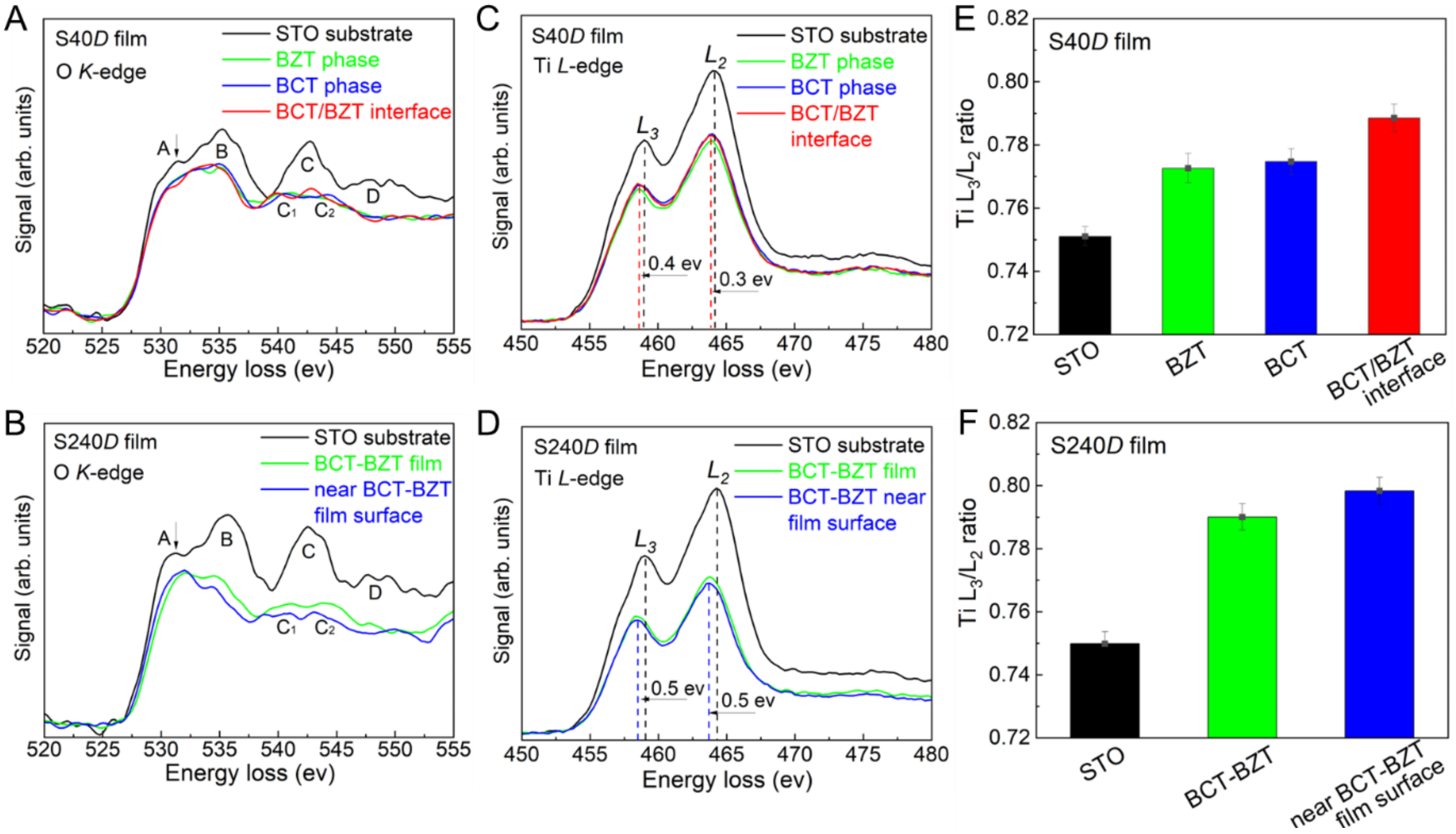

**Fig. 5 EELS spectroscopy results of S40*D* and S240*D* films at "optimal" property point.** (A, B) O *K*-edges, (C, D) Ti $L_{2,3}$-edges EELS spectra of S40*D* and S240*D* at different locations. (E, F) Ti $L_3/L_2$ while-line ratios of S40*D* and S240*D* films.

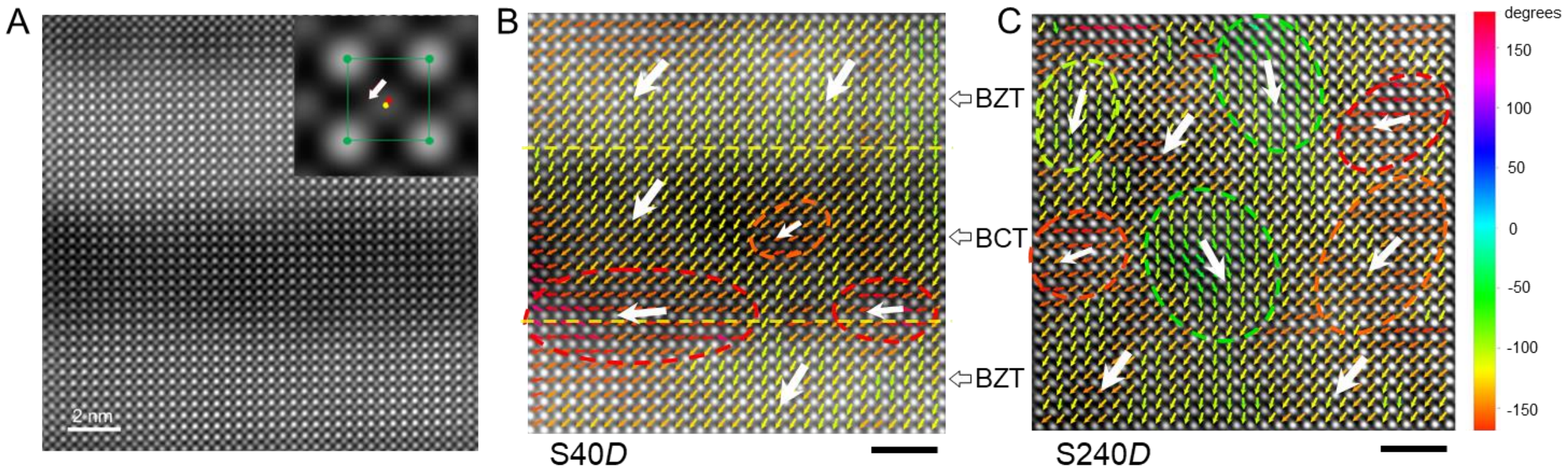

**Fig. 6 Polarization displacement vector maps.** (**A**) The HAADF-STEM image of the S40*C* BCT-BZT SL film. Inset is a zoom-in view of a single unit cell with the Ba (four corners) and Ti (center) atom positions denoted by green and yellow dots, respectively. The center-of-mass position of the four corner Ba atoms is denoted by the red cross. Displacement vector maps of (**B**) S40*D* and (**C**) S240*D* BCT-BZT thin films. The large white arrows denote the local mean shift of the Ti atoms. Scale bars: 2 nm.

Supplementary Materials for

# High-throughput combinatorial approach expedites the synthesis of a lead-free relaxor ferroelectric system

Di Zhang[1]*, Katherine J. Harmon[2], Michael J. Zachman[3], Ping Lu[4], Doyun Kim[5], Zhan Zhang[6], Nick Cucciniello[1], Reid Markland[1], Ken William Ssennyimba[1], Hua Zhou[6], Yue Cao[2], Matthew Brahlek[7], Hao Zheng[2,6], Matt Schneider[8], Alessandro Mazza[1,8], Zach Hughes[1], Chase Somodi[1], Benjamin Freiman[1], Sarah Pooley[1], Sundar Kunwar[1], Pinku Roy[1], Qing Tu[5], Rodney McCabe[8], Aiping Chen[1]*

*Corresponding authors. Email: dizhang@lanl.gov, apchen@lanl.gov

## Materials and Methods

**PLD growth of BCT-BZT thin films:** Pulsed laser deposition (PLD, KrF excimer laser, λ = 248 nm) was employed to grow BCT-BZT films on $SrRuO_3$ (SRO) buffered 10×5 mm $SrTiO_3$ (STO) (001) substrates purchased from CRYSTAL GmbH, Germany. BCT ($(Ba_{0.7}Ca_{0.3})TiO_3$) and BZT ($Ba(Zr_{0.2}Ti_{0.8})O_3$) targets were fabricated by a conventional solid-state reaction method, with mixing high purity $BaCO_3$ (99.95%), $CaCO_3$ (99.95%), $TiO_2$ (99.9%), and $BaZrO_3$ (99.5%) powders, followed by calcining and sintering process at 1100 $^oC$ in air for 12 h. Prior to the deposition, the chamber was pumped down to a base pressure of ~$2\times10^{-7}$ Torr. The laser fluence of 1.5 J/cm$^2$ and 2.0 J/cm$^2$ were used to grow the BCT-BZT films and SRO bottom electrode, respectively. The laser frequency of 2 Hz, oxygen partial pressure of 50 mTorr, and substrate temperature of 750 $^oC$ were maintained during the deposition process. The thickness of each individual layer in the BCT-BZT superlattice films was varied by tuning the deposition time for each layer, while the total film thicknesses were controlled by the total number of layers, *i.e.* 40, 80, and 240 in this work. After deposition, the films were cooled down at $10^o$ C/min under 500 Torr oxygen atmosphere to room temperature.

**X-ray diffraction (XRD) analysis:** X-ray reciprocal space mapping (RSM) was carried out on 40-layer and 240-layer BCT-BZT thin film samples at beamline 33-ID-D of the Advanced Photon Source (APS) using an X-ray photon energy of 15.5 keV (λ = 0.8 Å) and a beam focused to 20 um × 50 um (v × h). The out-of-plane lattice parameter and superlattice (SL) *d*-spacing of the films were determined by measuring RSMs in the vicinity of the substrate STO (002) Bragg reflection. Measurements were carried out in 200 um steps along the film composition gradient direction, spanning the full 25 mm length of each sample series in the case of the (002) RSMs. The film *d*-spacings as a function of position along the sample were evaluated from the peak L positions assuming Gaussian intensity distributions. In addition, specular crystal truncation rod (CTR) measurements were carried on the 80-layer BCT-BZT film series using an X-ray photon energy of 10 keV (λ = 1.24 Å) with the beam focused to 10 um × 20 um (v × h). Data were measured along the STO (002) CTR from L = 1.8 to L = 2.1 at 0.8 mm lateral increments along the film gradient direction, spanning the central 3.2 mm of each 5 mm sample of this series.

**Scanning transmission electron microscopy (STEM):** Cross-sectional TEM samples were prepared by a standard focused ion beam (FIB) lift-out process on a FEI Helios 600 dual beam SEM/FIB with $Ga^+$ ion source operated at 30 kV. The TEM lamellae were thinned to ~ 400-500 nm using a beam accelerating voltage of 30 kV and 0.5 nA followed by 100 pA beam current and were further thinned down to ~ 150-200 nm using beam parameters of 16 kV and 50 pA, followed by 27 pA. Final polishing was conducted using a 5 kV beam at 16 pA followed by 2 kV at 9 pA to remove the ion-damaged layer on the sample surface. Aberration-corrected STEM imaging was performed on a JEOL NEOARM in the Center for Nanophase Materials Sciences (CNMS) at Oak Ridge National Laboratory (ORNL) operated at 200 kV with a semi-convergence angle of ~ 28 mrad. Annular dark-field (ADF) images were acquired with a collection angle of 90-370 mrad. The SmartAlign plug-in (HREM Research Inc.) in Gatan DigitalMicrograph 3.5 (DM) software was used to perform linear and non-linear alignments of stacks of 30 individual images acquired consecutively in the same film region to correct linear- and non-linear image drift and scan distortions in atomic-resolution STEM images (*57,58*). Geometric phase analysis (GPA) maps were then derived from drift-corrected HAADF-STEM images to determine out-of-plane ($\varepsilon_{yy}$) and in-plane ($\varepsilon_{xx}$) strain using a GPA plug-in (HREM Research) in DM (*59*). To perform energy-dispersive X-ray spectroscopy (EDS), an FEI Titan™ G2 80-200 STEM with a Cs probe corrector and ChemiSTEM™ technology (X-FEG™ and SuperX™ EDS with four windowless silicon drift detectors), operated at 200 kV, was used. EDS spectral image data was acquired with the probe parallel to the BCT or BZT [100] zone axis with an electron probe of size <0.13 nm, a convergence angle of 17.1 mrad, and a current of ~ 100 pA. Electron energy-loss spectroscopy (EELS, Gatan 963) was used to probe the O *K*-edge fine structure of the films under similar optical conditions with an energy dispersion 0.25 eV/channel and instantaneous dwell time of 500 ms.

**Ferroelectric and dielectric properties measurement:** For the as-deposited BCT-BZT films, circular gold (Au) top electrodes with thickness about 100 nm were deposited by magnetron sputtering by using shadow masks with holes diameter of 150 μm. The real top electrodes areas are in general larger than the predesigned hole size on the mask. To remove the inaccuracy caused by non-uniformity of top Au electrode areas on the specimens, we measured the true areas of each electrode using scanning electron microscope (SEM) (Fig. S7). The ferroelectric polarization-voltage (*P-E*) hysteresis loops were measured using a Precision LC II Ferroelectric Tester (Radiant Technologies) at 10 Hz to 100 kHz, and the capacitance-voltage (*C-V*) and dielectric loss tangent results were measured using an Agilent E4980A LCR meter. Based on the capacitance equation $C = \varepsilon_0 \varepsilon_r A/d$, where $C$ is capacitance (F) of the device, $\varepsilon_0$ is the vacuum permittivity ($8.854\times10^{-12}$

F/m), $A$ and $d$ are electrode area (μm$^2$) and film thickness (nm), we can easily calculate the permittivity (dielectric constant) $\varepsilon_r$ values of our ferroelectric devices.

**Piezoelectric force microscopy (PFM):** All atomic force microscopy (AFM) topographical imaging and piezoresponse force microscopy (PFM) measurements were conducted with an MFP-3D Infinity AFM (Asylum Research, an Oxford instrument, CA). Prior to polarizing the sample and PFM measurements, we first identify a clean area and sweep the contaminants off the surface by multiple scans of contact mode imaging of the region (20×20 μm) at high contact forces using a stiff Si cantilever (RTESP, Bruker, spring constant ~ 40 N/m). The subsequent electrical polarization and PFM measurements were performed using a Pt coated tip (Multi75E-G, BudgetSensors). The deflection sensitivity of the cantilever was first calibrated by engaging the cantilever onto a clean silicon wafer. To ensure a saturation of the polarization and avoid any potential piezoresponse variations due to local differences in the polarization magnitude and orientation, we electrically polarize the cleaned region before performing the quantitative PFM measurements by applying +8V DC bias (tip grounded) in the white squares as schematically shown in Supplementary Fig. S9A. Then PFM point measurements were performed at the center of each polarized squares as described in (*60*). Briefly, an AC bias with an amplitude $V_{AC}$ was applied through the back electrode while the tip was grounded (Fig. 3F), and the cantilever response signal $A$ was swept across the first contact resonance of the AFM cantilever. The dynamics response of the AFM cantilever can be modeled as a damped simple harmonic oscillator (DSHO) and the obtained $A$ vs. frequency $\omega$ will be fitted to:

$$A(\omega) = \frac{A_{\text{piezo}}\omega_0^2}{\sqrt{(\omega^2-\omega_0^2)^2+(\omega\omega_0/Q)^2}} \qquad (1)$$

to extract the actual piezoresponse signal $A_{\text{piezo}}$ (Fig. S9B). $A_{\text{piezo}}$ was found to be a linear function of $V_{AC}$ (Fig. S9C), confirming the piezoelectric feature of our samples. The slope of $A_{\text{piezo}}$ *vs.* $V_{AC}$ is the measured effective piezoelectric coefficient $d_{33}$ of the sample.

**Polarization displacement vector map:** The determination of polar displacement vector map was performed based on the aberration-corrected HAADF-STEM images using local A- and B-site sublattice offset measurements (see Fig. 6A) (*61*). The rationale is that for many displacive perovskite ferroelectrics the electric dipole moment is manifested by relative shifts of the cations, as is the case for $BaTiO_3$, and this offset can be used to infer the full polarization. The HAADF images were chosen because of their high SNR ratio and sensitivity to chemical distribution, which is very helpful for us to accurately identify the cations positions within the lattice. Here, the atom position finding and 2D Gaussian refinement were completed with the Atomap Python package first (*62*), then the polarization vector maps were completed using the TopoTEM module (*55*) of the TEMUL Toolkit Python package.

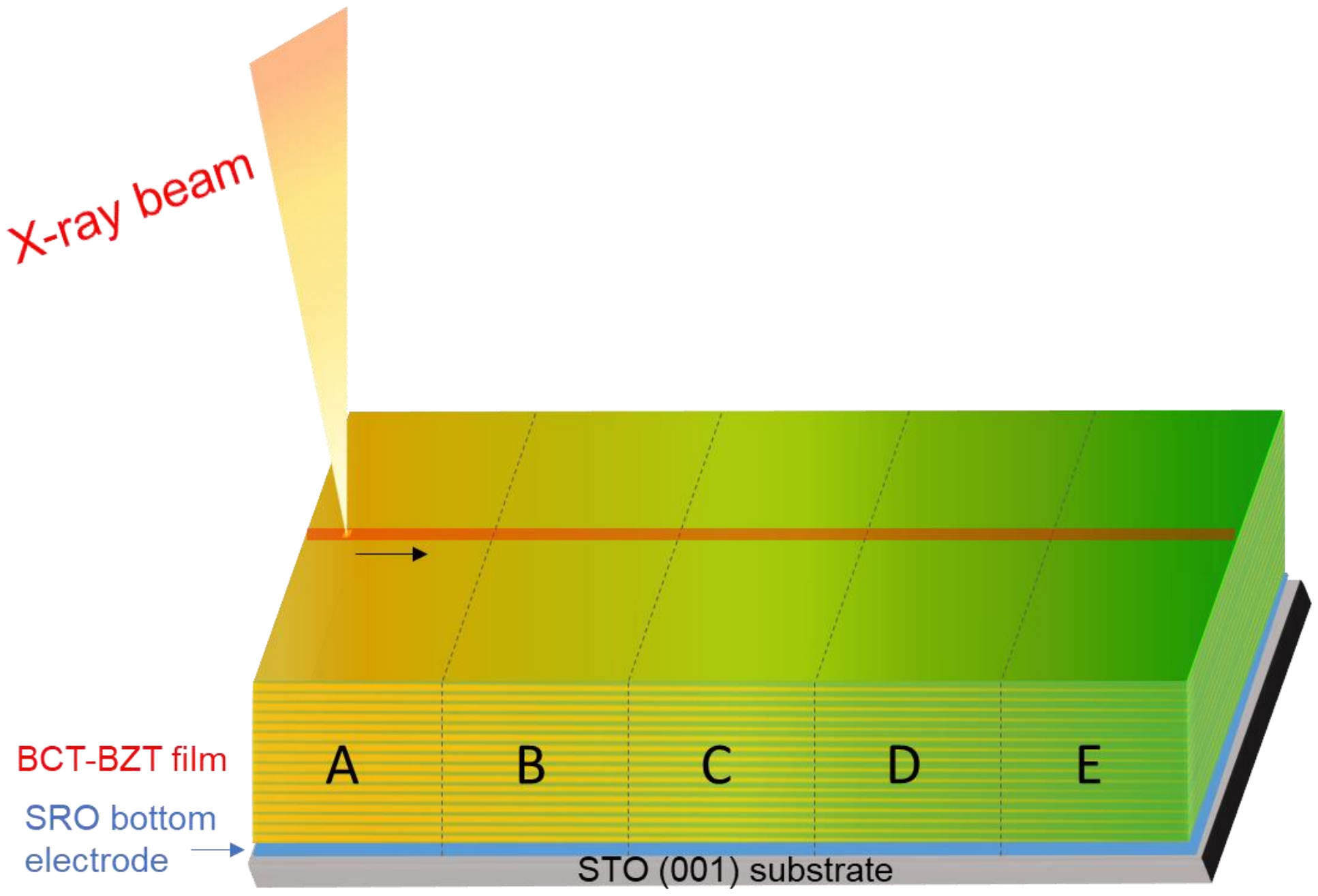

**Fig. S1** Schematic of X-ray reciprocal space mapping (RSM) experiment showing the beam scanning direction on the sample surface.

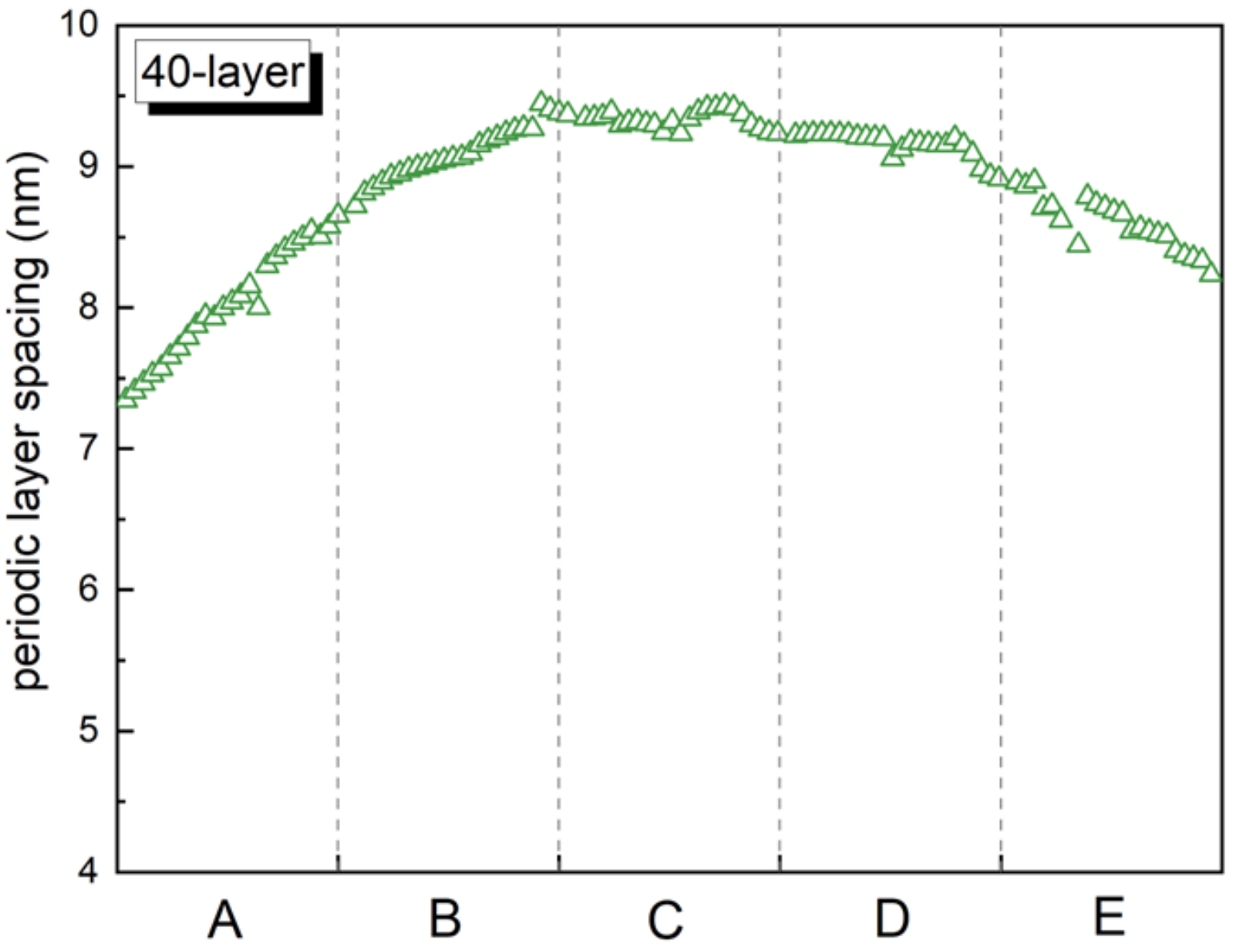

**Fig. S2 XRD results of the S40 BCT-BZT film**. Periodicity for the S40 SL film. The thickness of periodic layer $(BCT/BZT)_1$ were determined from the primary (002) and the -1 SL peaks positions.

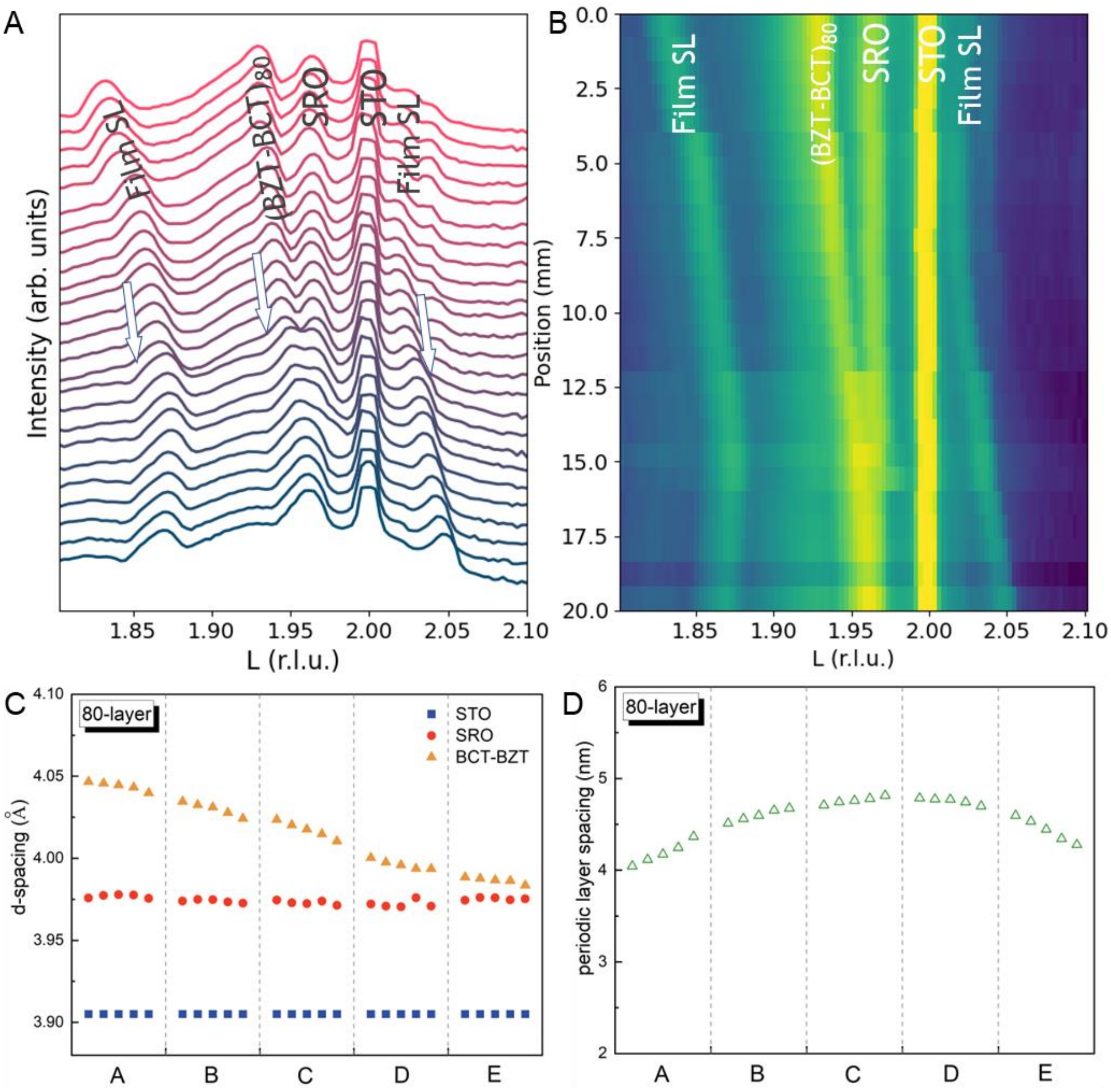

**Fig. S3 XRD results of the S80 BCT-BZT film.** (**A, B**) XRD intensities of the S80 BCT-BZT film. The curves were extracted from CTR data as function of reciprocal lattice units (r.l.u.) around STO (002) Bragg reflection. Calculated location-dependent (**C**) lattice *d*-spacing and (**D**) periodic layer spacing for the S80 film. The *d*-spacing of the films were determined by measuring RSMs in the vicinity of the substrate STO (002) Bragg reflection. The thicknesses of $(BCT/BZT)_1$ periodicity were determined from the film primary and the +1 satellite peaks positions.

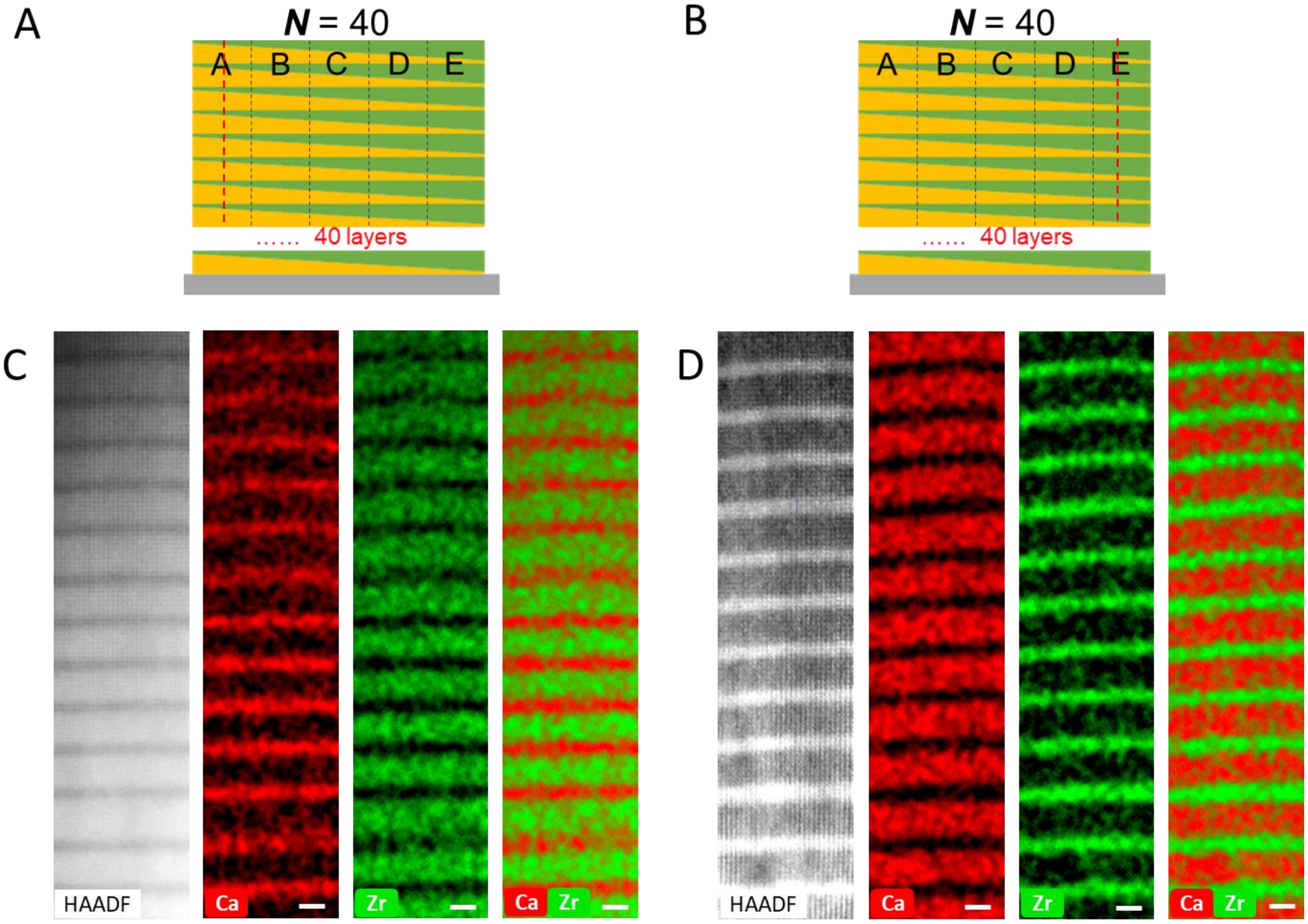

**Fig. S4 Microstructure analysis of S40 BCT-BZT film at location *A* and *E*.** (**A, B**) Schematics of the cross-sectional views with red dashed lines showing the investigated sample locations, and (**C, D**) HAADF-STEM and EDS elemental maps of Ca (red) and Zr (green) for S40*A* and S40*E*. Scale bars: 5 nm.

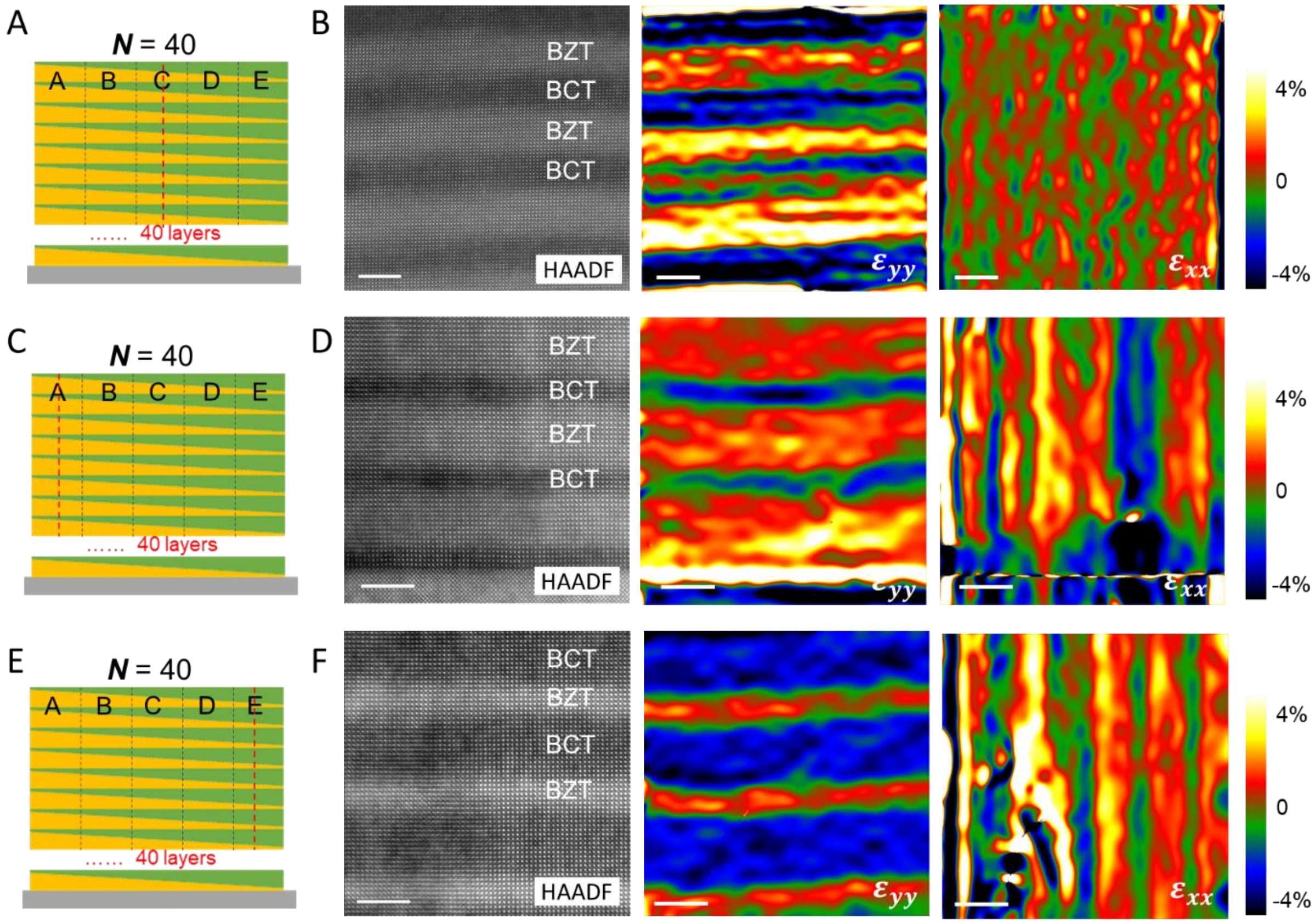

**Fig. S5 Microstructure and normal strain maps of S40 BCT-BZT films, S40*C*, S40*A* and S40*E*.** (**A**, **C, E**) Schematics of cross-sectional views with red lines indicating the sample location on the 40-layer BCT-BZT SL film, respectively, along with (**B, D, F**) HAADF-STEM images of the materials showing the (BCT/BZT) SLs and corresponding GPA analysis of out-of-plane ($\varepsilon_{yy}$) and in-plane ($\varepsilon_{xx}$) strain maps. In S40*C*, the BCT:BZT volume ratio is ~45:55. In S40*A*, the BCT:BZT volume ratio is ~25:75. In S40*E*, the BCT:BZT volume ratio is ~70:30. Scale bars: 5 nm.

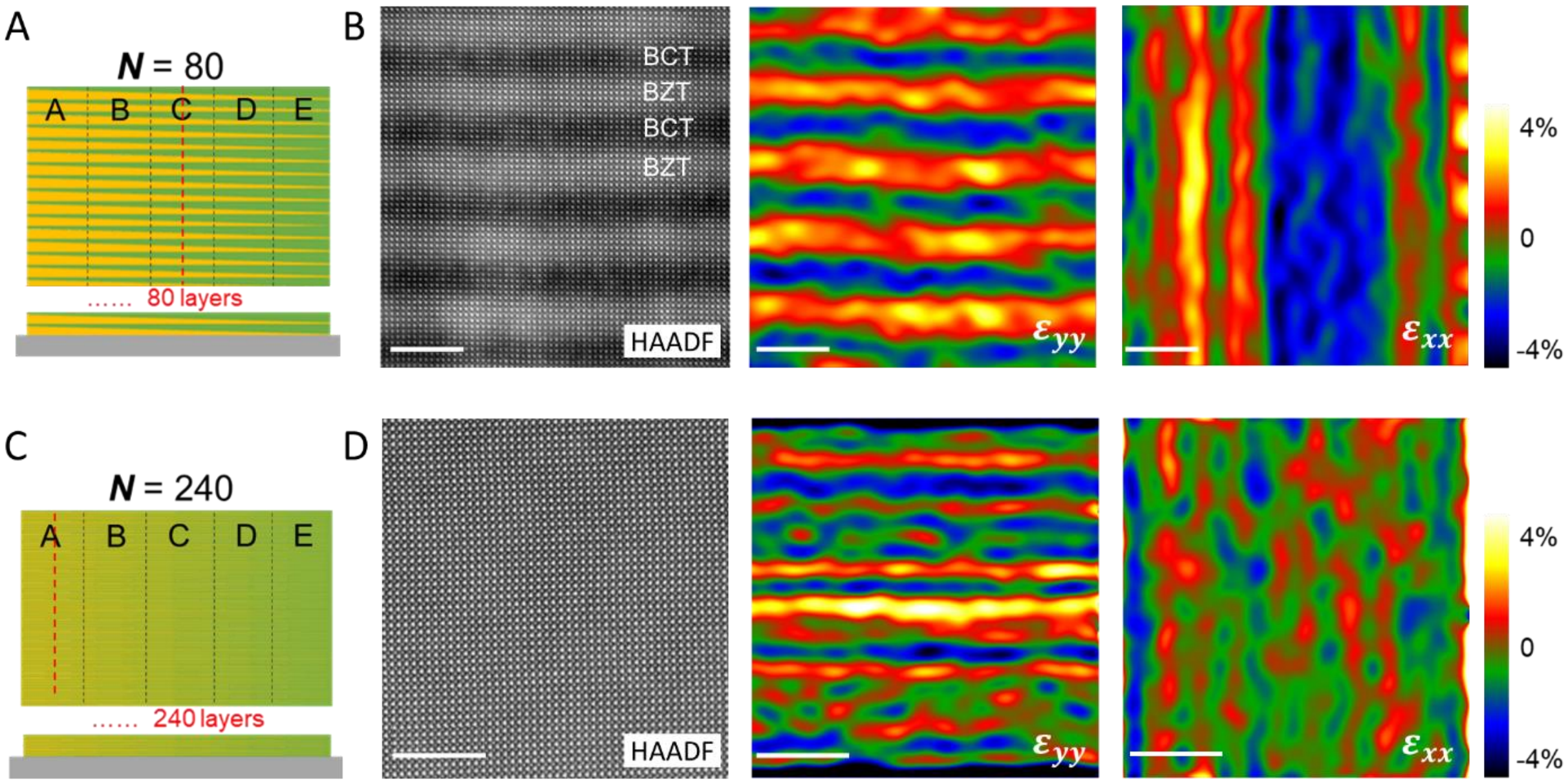

**Fig. S6 Microstructure and normal strain maps of S80*C* and S240*A* BCT-BZT films**. (**A**, **C**) Schematics of cross-sectional views with red lines indicating the sample location on the S80 and S240 BCT-BZT films, respectively, along with (**B, D**) HAADF-STEM images of the materials and corresponding GPA analysis of out-of-plane ($\varepsilon_{yy}$) and in-plane ($\varepsilon_{xx}$) strain maps. Scale bars: 5 nm.

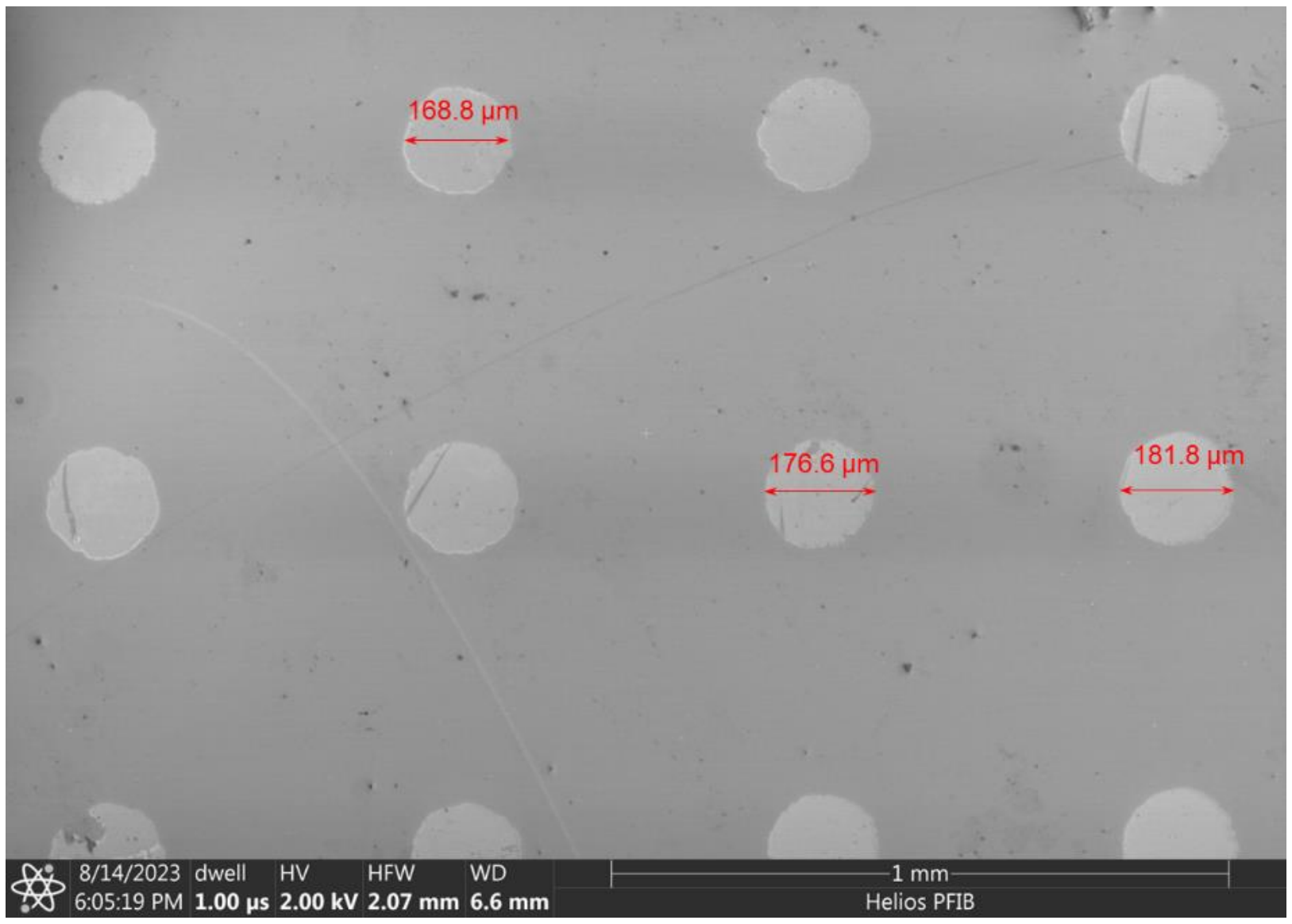

**Fig. S7 The top electrodes sizes measured by SEM**. The size of Au top electrodes measured by SEM. The actual electrode size is larger than the predefined hole sizes (150 µm) on the shadow mask.

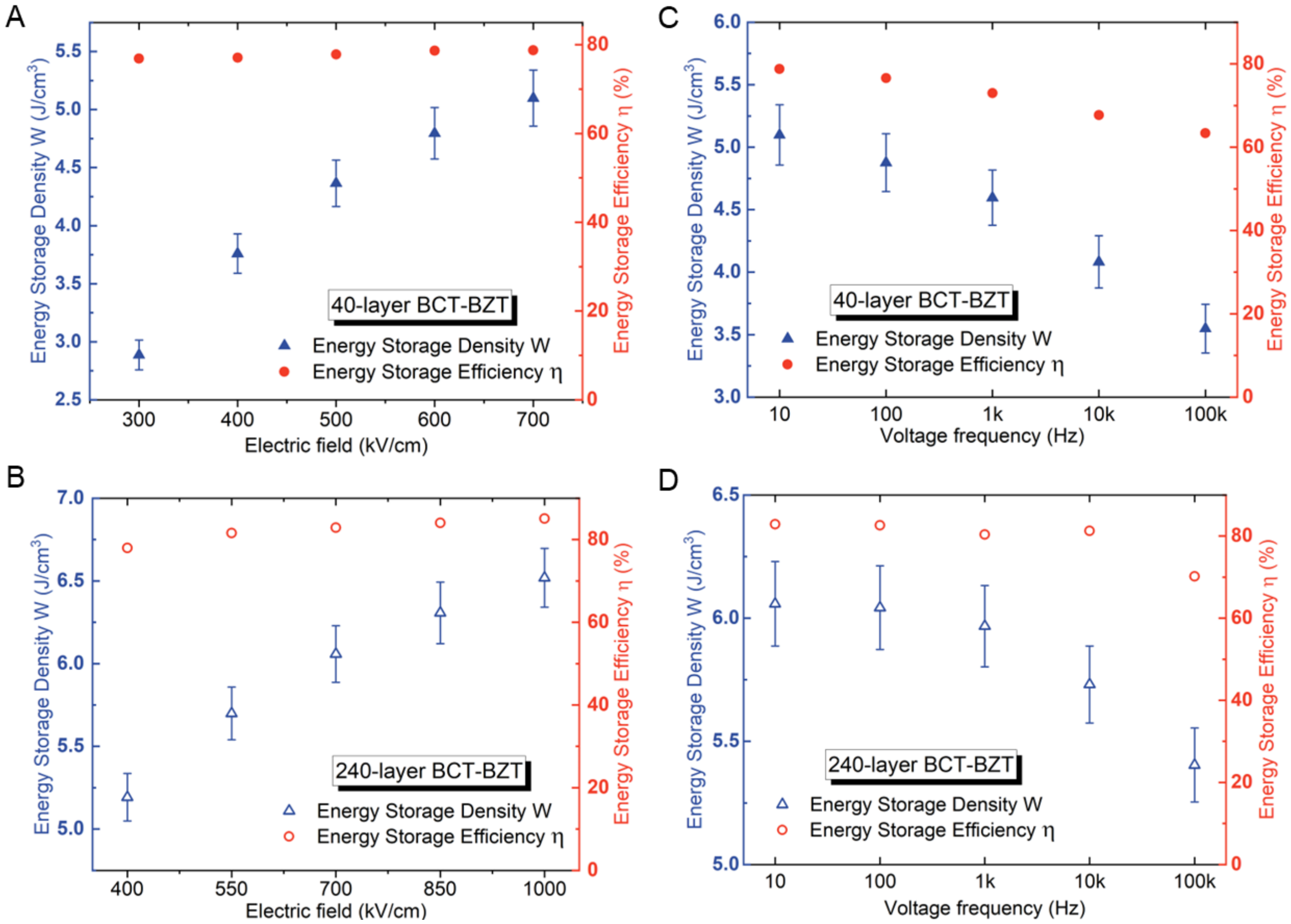

**Fig. S8 Energy storage density (*W*) and energy storage efficiency (*η*).** (**A**, **B**) Energy storage density(*W*) and energy storage efficiency (*η*) as functions of electric field (kV/cm) for S40 and S240 films. (**C**, **D**) Energy storage density(*W*) and energy storage efficiency (*η*) as functions of voltage frequency (Hz) for the S40 and S240 films.

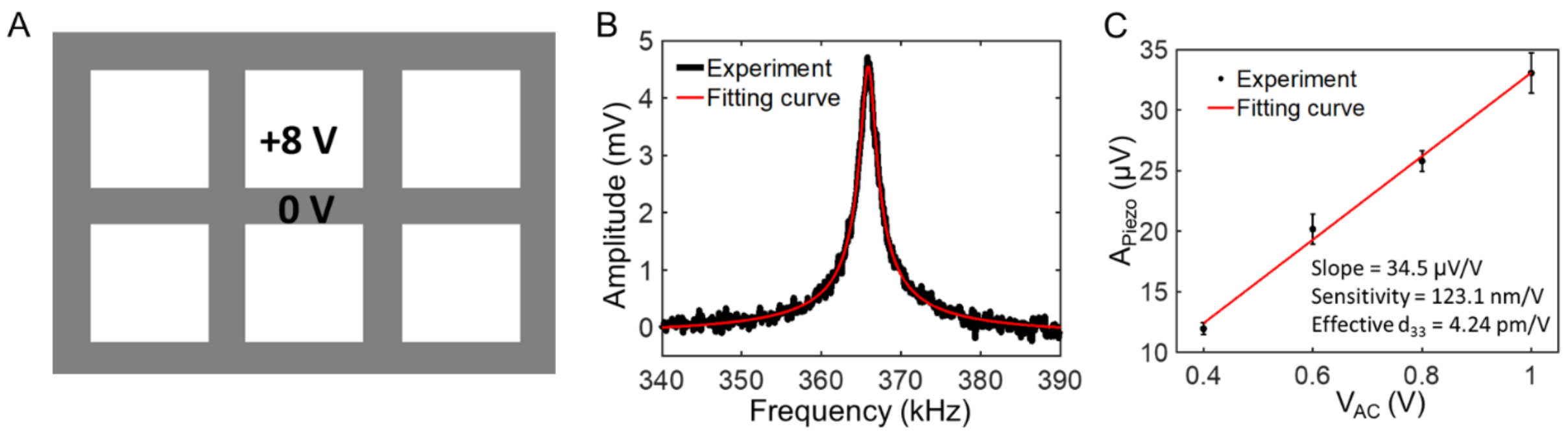

**Fig. S9 Piezoresponse force microscopy (PFM) measurement.** (**A**) Square patterns used to electrically polarize the sample for $d_{33}$ measurements. (**B**) Representative cantilever piezoresponse amplitude of S40*A* fitted with DSHO model (eq. (1)) and (**C**) the representative linear trend of $A_{piezo}$ as a function of driving AC voltage ($V_{AC}$). The error bar indicates the standard deviation of the mean.

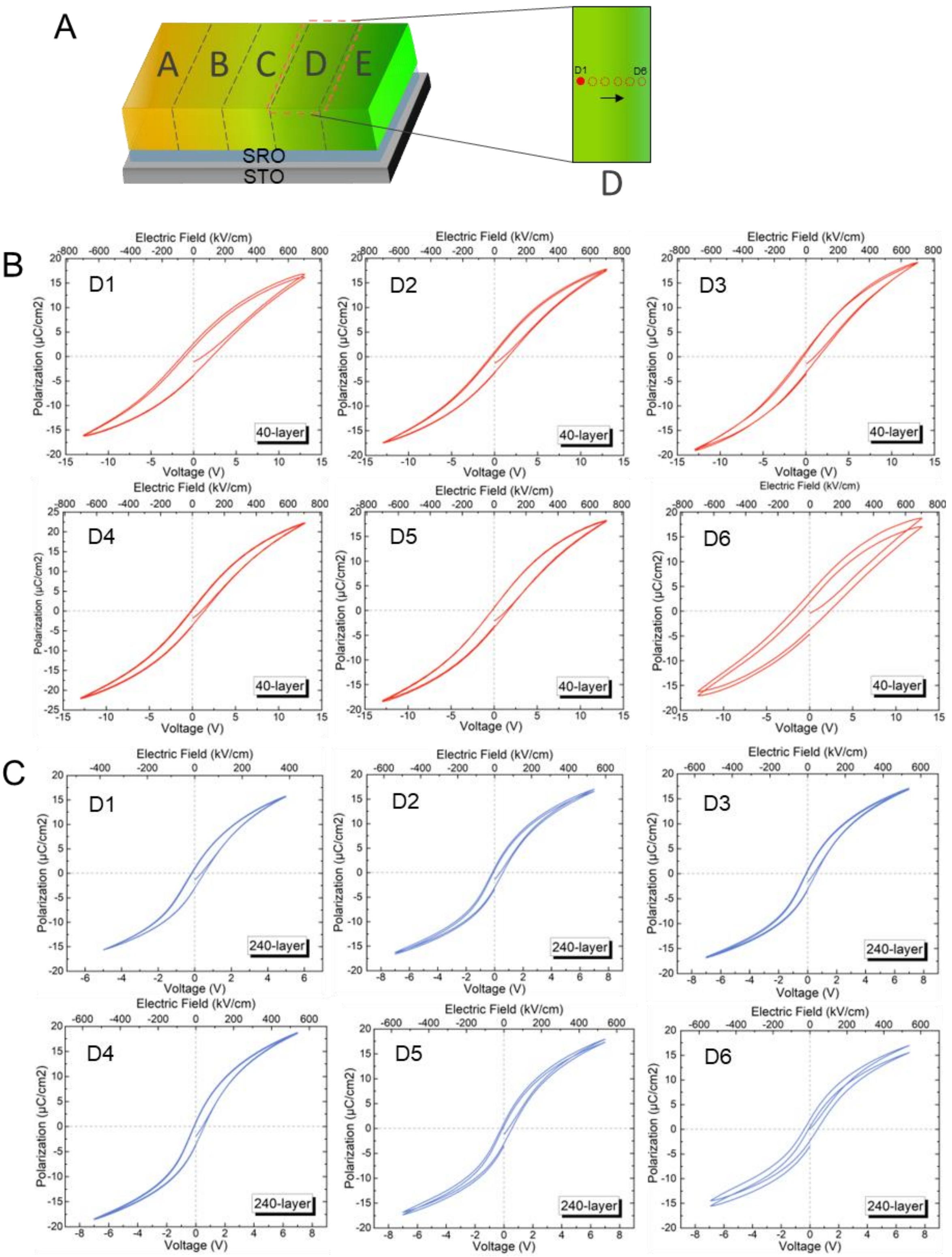

**Fig. S10 Ferroelectric measurement results**. (**A**) Schematic of the measured locations for S40 and S240 films. (**B**, **C**) Ferroelectric polarization *vs.* electric field (voltage) *P*-*E* hysteresis loops of the measured datapoints at sample *D* for S40 and S240 films.

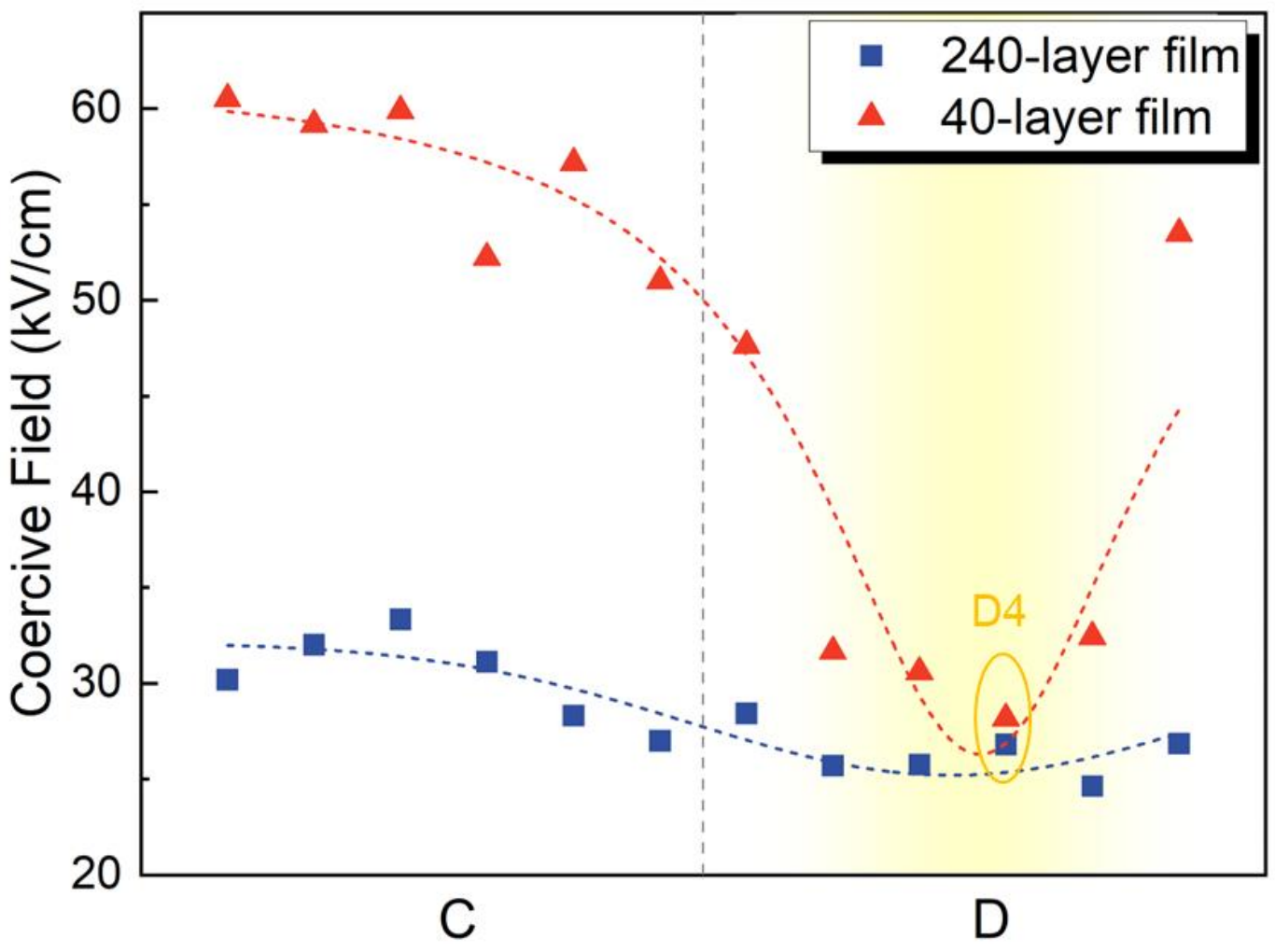

**Fig. S11 Coercive fields of the BCT-BZT films**. Coercive fields ($H_c$) of S40 and S240 BCT-BZT thin films at sample *C* and *D*. It shows that the smallest $H_c$ values obtained at *D*4 position for both films.

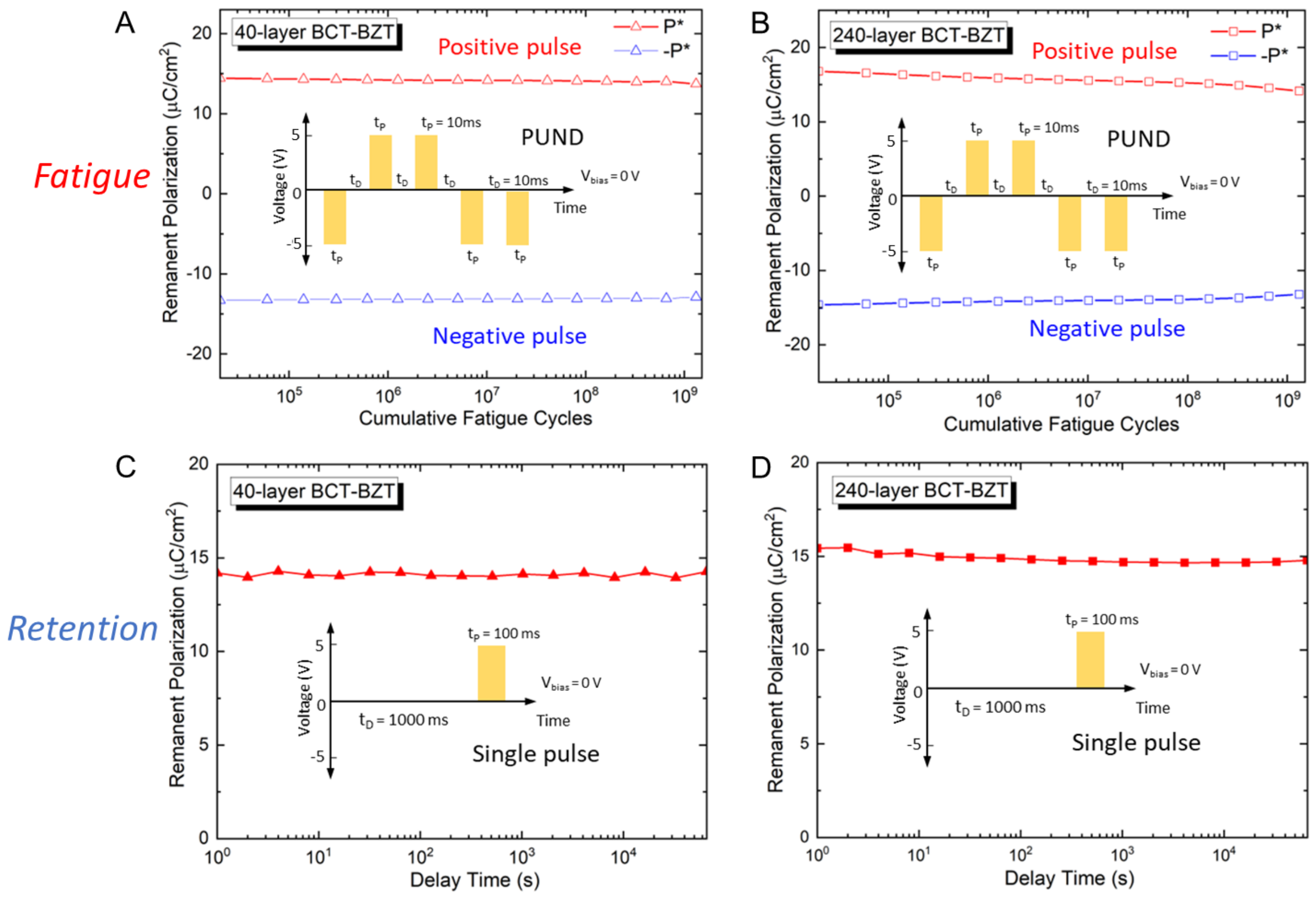

**Fig. S12 Fatigue and Retention measurements**. Fatigue (**A**, **B**) and Retention (**C**, **D**) measurements results of S40 and S240 BCT-BZT films. Insets show the pulse sequences used to probe remanent polarization in fatigue and retention measurements, respectively. "PUND" denotes the Positive-Up-Negative-Down five-pulse measurement termed by Radiant Technologies.

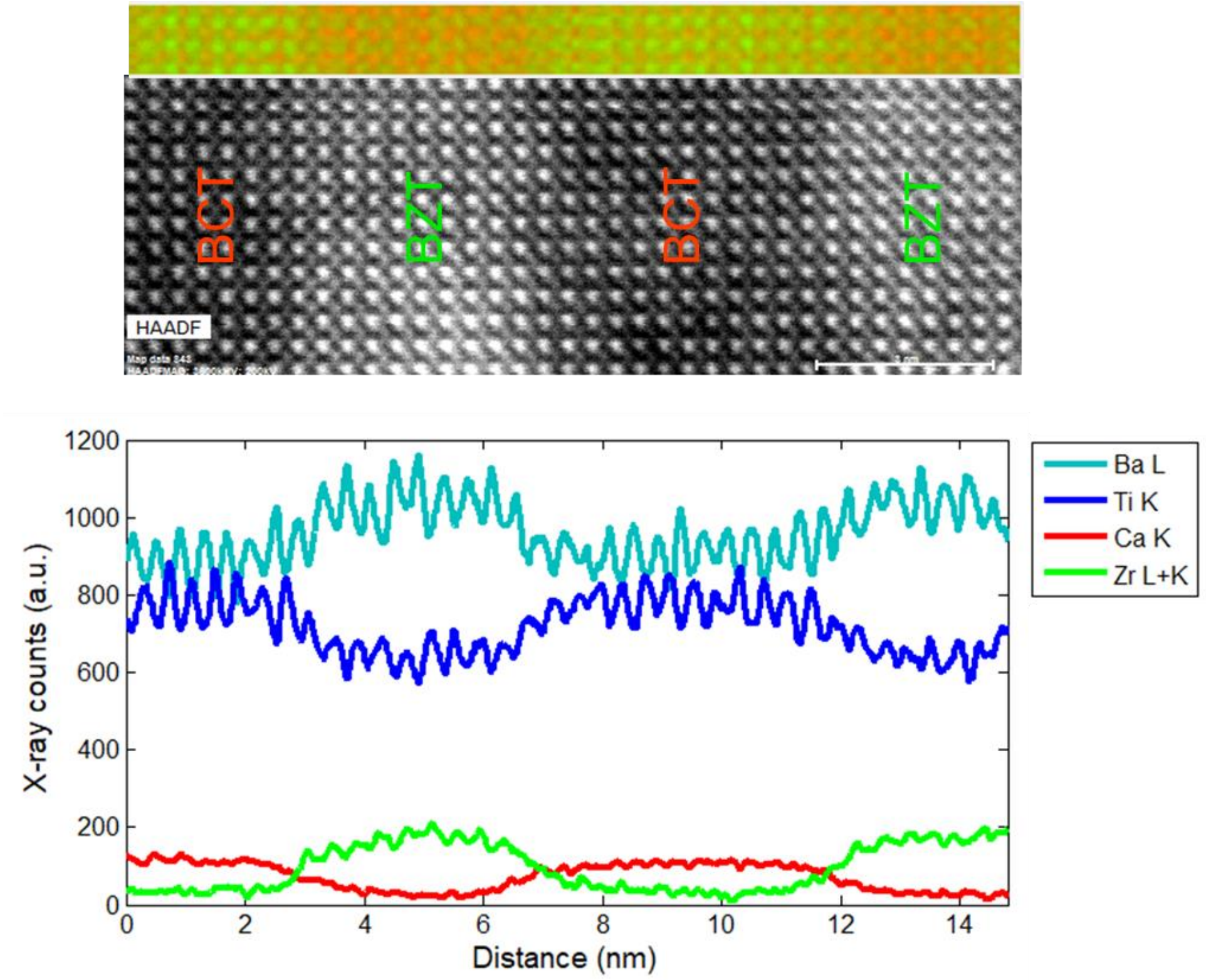

**Fig. S13 EDS intensity line profile of S40D**. The EDS intensity line profiles of individual elements (Ba, Ti, Ca, Zr) within S40*D* BCT-BZT SL film at "optimal property" location. Scale bar: 3 nm.

**Table S1 Elements quantification table of S40*D* BCT-BZT SL film.**

| **40-layer BCT-BZT at "optimal property point"** | | | |
|---|---|---|---|
| Element | Series | weight percent [wt.%] | Atomic percent [at.%] |
| Calcium (Ca) | K-series | 4.00 | 7.14 |
| Zirconium (Zr) | K-series | 8.52 | 6.69 |
| Barium (Ba) | L-series | 45.84 | 23.90 |
| Titanium (Ti) | K-series | 41.64 | 62.27 |
| | Total: | 100 | 100 |
| The atomic percentage ratio of Ca/Zr is 7.14/6.69 = 1.07 | | | |

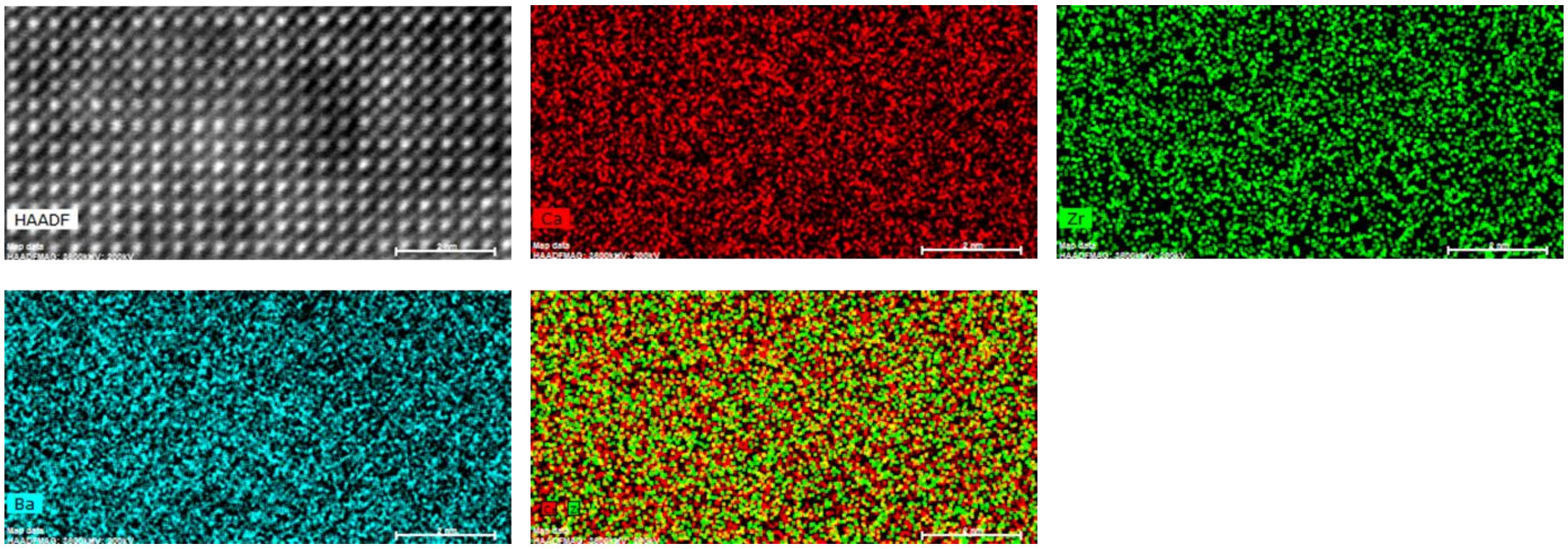

**Fig. S14 EDS elemental maps of S240D BCT-BZT film**. EDS elemental maps of Ca, Zr, Ba and composite map of S240*D* BCT-BZT film at "optimal property" location. Scale bars: 2 nm.

**Table S2 Elements composition table of S240*D* BCT-BZT film.**

| **240-layer BCT-BZT at "optimal property point"** | | | |
|---|---|---|---|
| Element | Series | weight percent [wt.%] | Atomic percent [at.%] |
| Calcium (Ca) | K-series | 4.19 | 7.52 |
| Zirconium (Zr) | K-series | 8.62 | 6.80 |
| Barium (Ba) | L-series | 46.28 | 24.24 |
| Titanium (Ti) | K-series | 40.91 | 61.44 |
| | Total: | 100 | 100 |
| The atomic percentage ratio of Ca/Zr is 7.52/6.80 = 1.10 | | | |

**Note:** The EDS quantification results shown in Table S1 and S2 are standardless. Due to the strong overlap between Ba and Ti peaks, the deconvolution results based on Ba and Ti peak intensities are not reliable. For Ca and Zr peaks, however, since there is no peak overlapping which can cause strong inaccuracy, the Ca/Zr atomic ratios are more reliable and trustworthy.

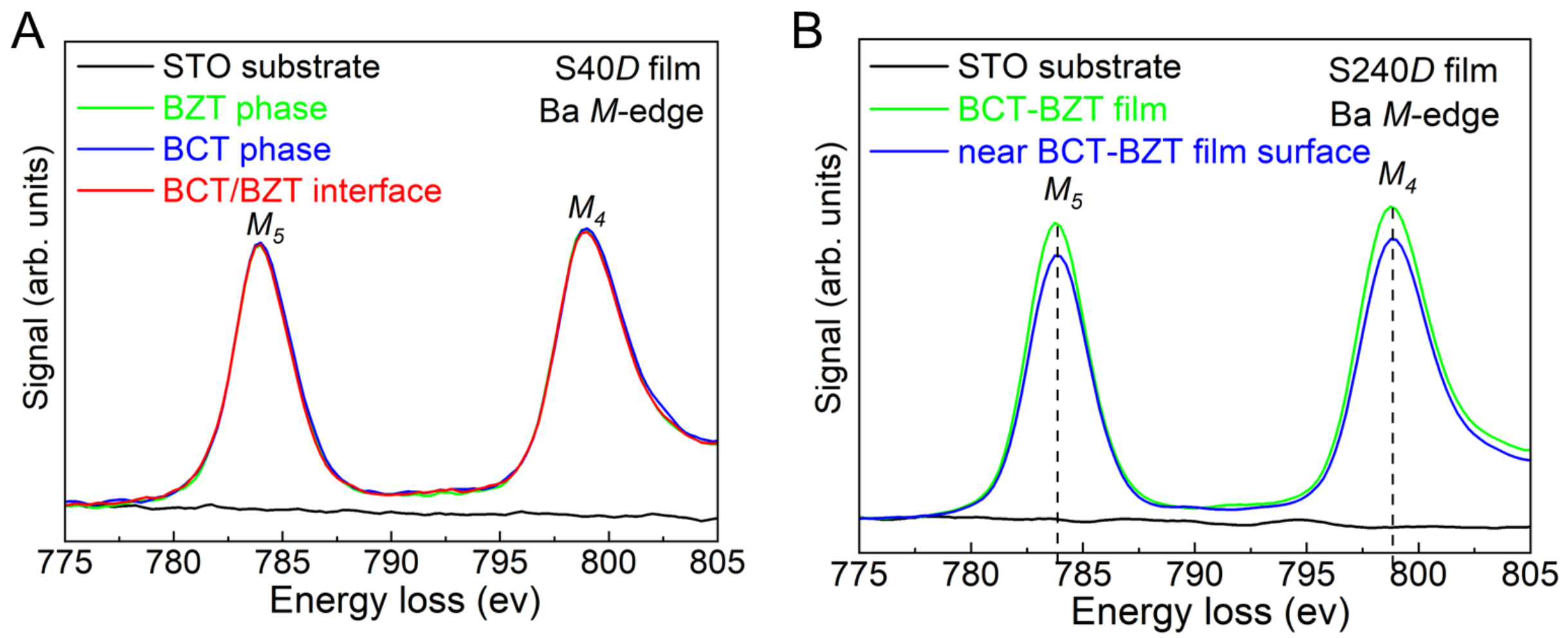

**Fig. S15 Ba $M_{4,5}$-edge EELS spectra for S40*D* and S240*D* BCT-BZT films**. Ba $M_{4,5}$-edges EELS spectra of the (**A**) S40*D* and (**B**) S240*D* extracted from different film locations.

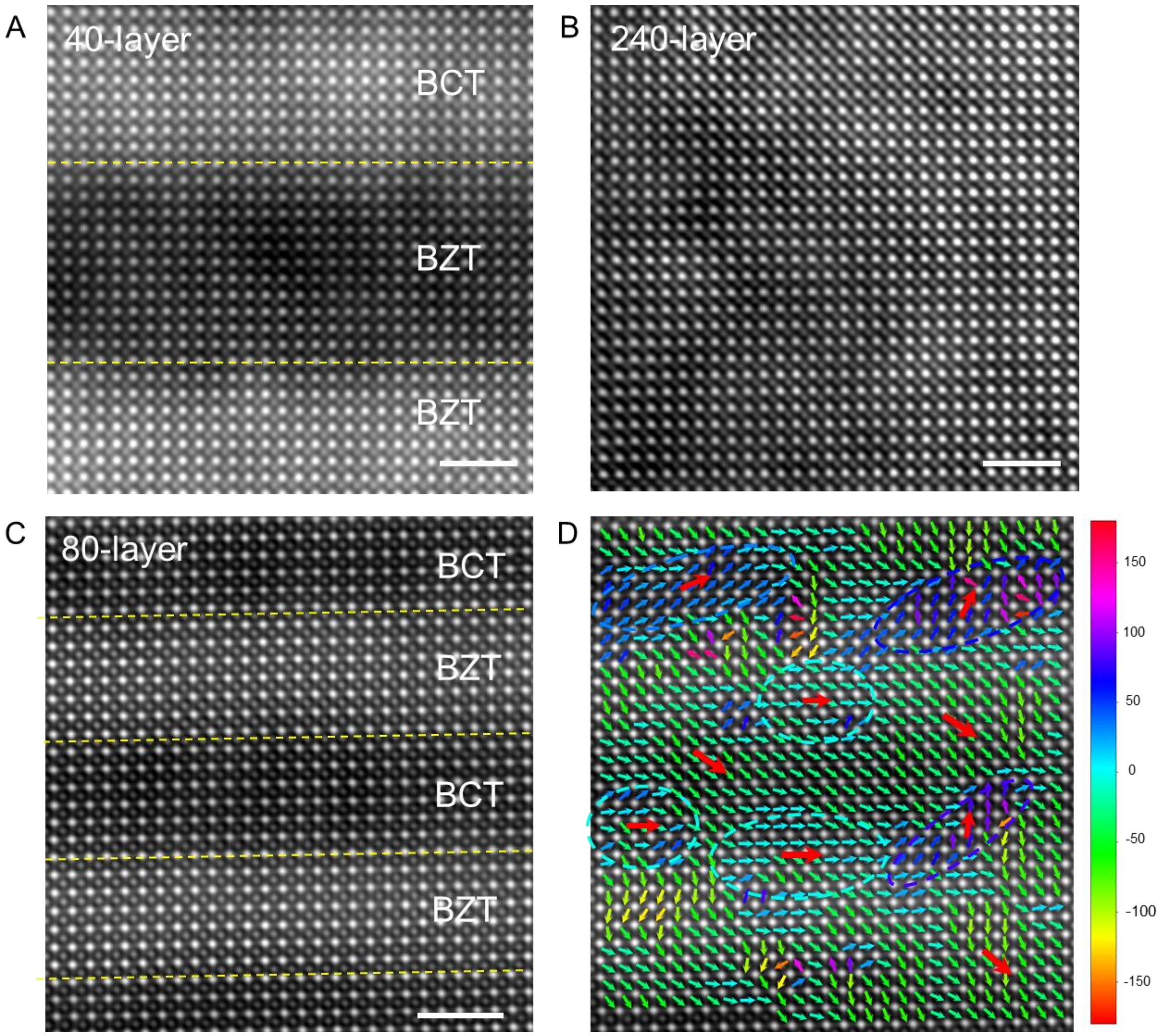

**Fig. S16 HAADF-STEM images and polarization displacement vector map**. HAADF-STEM images of (A) S40*D*, (B) S240*D*, and (C) S80*C* BCT-BZT films. (D) polarization displacement vector map for the S80C SL film. Scale bar: 2 nm.